\documentclass{elsarticle}

\usepackage[reviewcopy]{adndt}%{adndt} at NSCL, {../adndt} at Gross
\usepackage{longtable}

\usepackage{amsmath}

\usepackage{amssymb}

\biboptions{square,sort&compress}
\bibpunct[]{[}{]}{,}{n}{}{;}
\begin{document}

\begin{frontmatter}

\journal{Atomic Data and Nuclear Data Tables}

%% Author, fill in article title here

\title{Discovery of Samarium, Europium, Gadolinium, and Terbium Isotopes}

%% Fill in author list here
\author{E. May}
\author{M. Thoennessen\corref{cor1}}\ead{thoennessen@nscl.msu.edu}

 \cortext[cor1]{Corresponding author.}

 \address{National Superconducting Cyclotron Laboratory and \\ Department of Physics and Astronomy, Michigan State University, \\ East Lansing, MI 48824, USA}

\begin{abstract}
Currently, thirty-four samarium, thirty-four europium, thirty-one gadolinium, and thirty-one terbium isotopes have been observed and the discovery of these isotopes is discussed here. For each isotope a brief synopsis of the first refereed publication, including the production and identification method, is presented.
\end{abstract}

\end{frontmatter}

%%% Keywords and subject classification are not used in ADNDT
%%%\begin{keywords}
%%%Insert list of keywords here.
%%%\end{keywords}

%%%\begin{subject}[Insert header for classifications]
%%%Use only if your journal has a subject classification requirement
%%%\end{subject}

%%% The table of contents should start a new page. This command will
%%% automatically pull all the unstarred \section, \subsection and
%%% \subsubsection titles into the Contents. Starred versions need to be
%%% done manually using
%%%      \addcontentsline{toc}{[[sub]sub]section}{Section title}
%%% at the correct place. Examples are given below.

%%% The lists of data figures and data tables are created automatically
%%% by the \listofDfigures and \listofDtables commands.

\newpage
\tableofcontents
%%\listofDfigures
\listofDtables

\vskip5pc

\section{Introduction}\label{s:intro}

The discovery of samarium, europium, gadolinium, and terbium isotopes is discussed as part of the series summarizing the discovery of isotopes, beginning with the cerium isotopes in 2009 \cite{2009Gin01}. Guidelines for assigning credit for discovery are (1) clear identification, either through decay-curves and relationships to other known isotopes, particle or $\gamma$-ray spectra, or unique mass and Z-identification, and (2) publication of the discovery in a refereed journal. The authors and year of the first publication, the laboratory where the isotopes were produced as well as the production and identification methods are discussed. When appropriate, references to conference proceedings, internal reports, and theses are included. When a discovery includes a half-life measurement the measured value is compared to the currently adopted value taken from the NUBASE evaluation \cite{2003Aud01} which is based on the ENSDF database \cite{2008ENS01}. In cases where the reported half-life differed significantly from the adopted half-life (up to approximately a factor of two), we searched the subsequent literature for indications that the measurement was erroneous. If that was not the case we credited the authors with the discovery in spite of the inaccurate half-life. All reported half-lives inconsistent with the presently adopted half-life for the ground state were compared to isomers half-lives and accepted as discoveries if appropriate following the criterium described above.

The first criterium excludes measurements of half-lives of a given element without mass identification. This affects mostly isotopes first observed in fission where decay curves of chemically separated elements were measured without the capability to determine their mass. Also the four-parameter measurements (see, for example, Ref. \cite{1970Joh01}) were, in general, not considered because the mass identification was only $\pm$1 mass unit.

The second criterium affects especially the isotopes studied within the Manhattan Project. Although an overview of the results was published in 1946 \cite{1946TPP01}, most of the papers were only published in the Plutonium Project Records of the Manhattan Project Technical Series, Vol. 9A, ``Radiochemistry and the Fission Products,'' in three books by Wiley in 1951 \cite{1951Cor01}. We considered this first unclassified publication to be equivalent to a refereed paper.

The initial literature search was performed using the databases ENSDF \cite{2008ENS01} and NSR \cite{2008NSR01} of the National Nuclear Data Center at Brookhaven National Laboratory. These databases are complete and reliable back to the early 1960's. For earlier references, several editions of the Table of Isotopes were used \cite{1940Liv01,1944Sea01,1948Sea01,1953Hol02,1958Str01,1967Led01}. A good reference for the discovery of the stable isotopes was the second edition of Aston's book ``Mass Spectra and Isotopes'' \cite{1942Ast01}.

\section{Discovery of $^{129-162}$Sm}

Thiry-four samarium isotopes from A = 129--162 have been discovered so far; these include 7 stable, 17 neutron-deficient and 10 neutron-rich isotopes. According to the HFB-14 model \cite{2007Gor01}, on the neutron-rich side the bound isotopes should reach at least up to $^{202}$Sm while on the neutron deficient side four more isotopes should be particle stable ($^{125-128}$Sm). Six additional isotopes ($^{119-124}$Sm) could still have half-lives longer than 10$^{-9}$~ns \cite{2004Tho01}. Thus, about 50 isotopes have yet to be discovered corresponding to 60\% of all possible samarium isotopes.

\begin{figure}
	\centering
	\includegraphics[scale=.7]{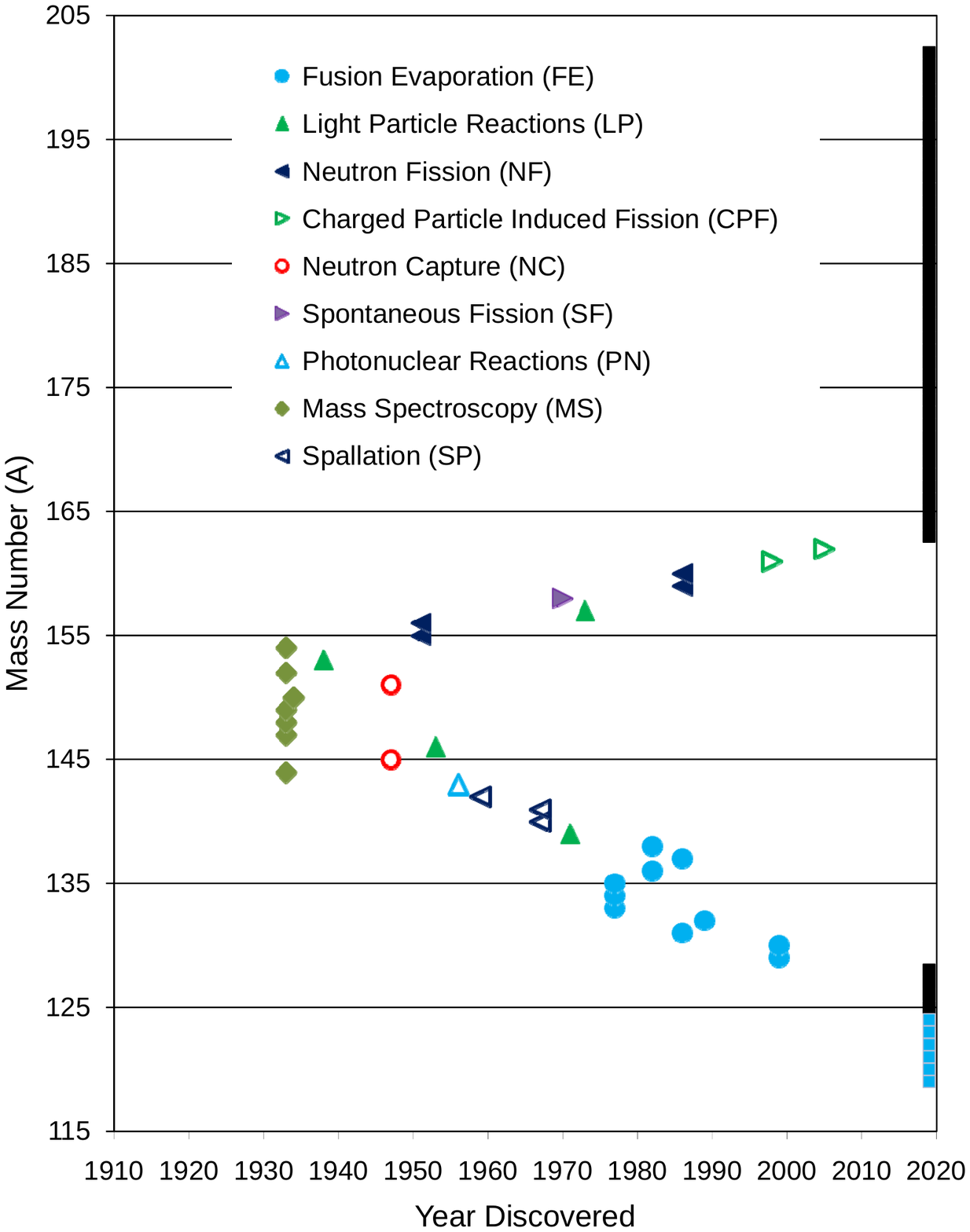}
	\caption{Samarium isotopes as a function of time when they were discovered. The different production methods are indicated. The solid black squares on the right hand side of the plot are isotopes predicted to be bound by the HFB-14 model. On the proton-rich side the light blue squares correspond to unbound isotopes predicted to have lifetimes larger than $\sim 10^{-9}$~s.}
\label{f:year-sm}
\end{figure}

Figure \ref{f:year-sm} summarizes the year of discovery for all samarium isotopes identified by the method of discovery. The range of isotopes predicted to exist is indicated on the right side of the figure. The radioactive samarium isotopes were produced using fusion evaporation reactions (FE), light-particle reactions (LP), neutron induced fission (NF), charged-particle induced fission (CPF), spontaneous fission (SF), neutron capture (NC), photo-nuclear reactions (PN), and spallation (SP). The stable isotope was identified using mass spectroscopy (MS). Light particles also include neutrons produced by accelerators. The discovery of each samarium isotope is discussed in detail and a summary is presented in Table 1.

\subsection*{$^{129}$Sm}\vspace{0.0cm}
Xu et al. first identified $^{129}$Sm in 1999 and reported the results in ``New $\beta$-delayed proton precursors in the rare-earth region near the proton drip line'' \cite{1999Xu01}. A 165 MeV $^{36}$Ar beam was accelerated with the Lanzhou sector-focused cyclotron and bombarded an enriched $^{96}$Ru target. Proton-$\gamma$ coincidences were measured in combination with a He-jet type transport system. ``A 134-keV $\gamma$ line found in the proton coincident $\gamma$($x$)-ray spectrum in the $^{36}$Ar+$^{96}$Ru reaction was assigned to the $\gamma$-ray transition between the lowest-energy 2$^{+}$ state and 0$^{+}$ ground state in the `daughter' nucleus $^{128}$Nd of the $\beta p$ precursor $^{129}$Sm.'' The observed half-life of 0.55(10)~s is currently the accepted value for $^{129}$Sm.

\subsection*{$^{130}$Sm}\vspace{0.0cm}
Sonzogni et al. reported the observation of $^{130}$Sm in the 1999 paper ``Fine structure in the decay of highly deformed proton emitter $^{131}$Eu'' \cite{1999Son01}. A $^{58}$Ni target was bombarded with a 402~MeV $^{78}$Kr beam from the ATLAS accelerator facility and $^{140}$Sm was populated following the proton decay of the fusion-evaporation product $^{131}$Eu. The recoil products were separated with the Argonne Fragment Mass Analyzer and stopped in a double-sided silicon strip detector. ``In addition to the previously observed ground-state line, measured here with a proton energy
of 932(7)~keV, a second proton peak with energy 811(7)~keV was observed. We interpret this line as proton decay from the $^{131}$Eu ground state to the first excited 2$^+$ state of the daughter nucleus $^{130}$Sm.''

\subsection*{$^{131}$Sm}\vspace{0.0cm}
The first observation of $^{131}$Sm was reported by Wilmarth et al. in their 1986 paper entitled ``Beta-delayed proton emission in the lanthanide region'' \cite{1986Wil01}. A 208 MeV $^{40}$Ca beam from the Berkeley Super HILAC bombarded a $^{96}$Ru target and $^{131}$Sm was produced in the fusion-evaporation reaction $^{96}$Ru($^{40}$Ca,2p3n). Beta-delayed particles, X-rays and $\gamma$-rays were measured following mass separation with the on-line isotope separator OASIS. ``A 1.2$\pm$0.2 s $\beta$-delayed proton activity coincident with Pm K x-rays identified the new isotope $^{131}$Sm'' The quoted half-life is currently used for the accepted value.

\subsection*{$^{132}$Sm}\vspace{0.0cm}
Wadsworth et al. discovered $^{132}$Sm in 1989 as reported in ``Yrast band spectroscopy of $^{132}$Sm and $^{130}$Nd'' \cite{1989Wad01}. A $^{96}$Ru target was bombarded with a 195~MeV $^{40}$Ca beam to produce $^{132}$Sm in the fusion-evaporation reaction $^{96}Ru$($^{40}$Ca,2p2n). Reaction products were separated with the Daresbury recoil separator and $\gamma$-rays and neutrons were measured with 10 bismuth germanate suppressed germanium detectors and five NE213 liquid scintillators, respectively. ``Transitions in $^{132}$Sm were assigned to $^{132}$Sm by comparison of gamma-ray spectra gated by mass 132 recoils and a single coincident neutron. The identification was helped by a further comparison with a 2n-$\gamma$ spectrum. This latter spectrum made it possible to distinguish between gamma-rays from $^{132}$Pr(3pn channel) and $^{132}$Sm(2p2n).''

\subsection*{$^{133-135}$Sm}\vspace{0.0cm}
The first identification of $^{133}$Sm, $^{134}$Sm, and $^{135}$Sm, was reported in 1977 by Bogdanov et al. in ``New neutron-deficient isotopes of barium and rare-earth elements'' \cite{1977Bog01}. The Dubna U-300 Heavy Ion Cyclotron accelerated a $^{32}$S beam which bombarded enriched targets of $^{102}$Pd and $^{106}$Cd. The isotopes were identified with the BEMS-2 isotope separator. ``By using the BEMS-2 isotope separator with a heavy-ion beam, we succeeded in producing 19 new isotopes with mass numbers ranging from 117 to 138. Five of these ($^{117}$Ba, $^{129, 131}$Nd and$^{133,135}$Sm) turned out to be delayed proton emitters.'' The reported half-lives of 3.2(4)~s for $^{133}$Sm, 12(3)~s for $^{134}$Sm, and 10(2)~s for $^{135}$Sm are close to the accepted values of 2.9(17)~s, 10(1)~s, and 10.3(5)~s, respectively. The half-life of 32$\pm$0.4~s mentioned for $^{133}$Sm in the table was apparently a typographical error.

\subsection*{$^{136}$Sm}\vspace{0.0cm}
``Very neutron deficient isotopes of samarium and europium'' by Nowicki et al. reported the discovery of $^{136}$Sm \cite{1982Now01}. An enriched $^{112}$Sn target was bombarded with a 190 MeV $^{32}$S beam from the Dubna U-300 cyclotron. The reaction products were identified with the on-line BEMS-2 mass separator and by measuring X- and $\gamma$-rays. ``Three isotopes: $^{136}$Sm, $^{137}$Eu, and $^{138}$Eu (with half-lives 40$\pm$5 s, 11$\pm$2 s and 12$\pm$2 s respectively) were observed for the first time''. The quoted half-life of 40(5)~s in agreement with the currently accepted value of 47(2)~s. The paper was submitted in October 1980 and between submission and publication two independent measurements of the $^{136}$Sm half-life were reported \cite{1981Kir01,1982Alk01}.

\subsection*{$^{137}$Sm}\vspace{0.0cm}
Redon et al. described the first observation of $^{137}$Sm in the 1986 paper ``Exotic neutron-deficient nuclei near N=82'' \cite{1986Red01}. Enriched $^{106}$Cd targets were bombarded with a 191 MeV $^{35}$Cl beam from the Grenoble SARA accelerator and $^{137}$Sm was formed in the fusion evaporation residue reaction $^{106}$Cd($^{35}$Cl,3p1n). The residues were separated with an on-line mass separator and a He-jet system. X-ray and $\gamma$-ray spectra were measured. ``A number of $\gamma$-rays with T$_{1/2}$=45$\pm$4 s were observed in the $^{35}$Cl + $^{106}$Cd reaction products. That confirms the result of Westgaard et al. on a (44$\pm$8)s activity obtained with a 600 Mev proton beam bombarding a molten Gd-La target.'' This half-life agrees with the currently accepted value of 45(1)~s. The previous work by Westgaard et al. mentioned in the quote was only published in a conference proceeding \cite{1973Wes01}.

\subsection*{$^{138}$Sm}\vspace{0.0cm}
``Very neutron deficient isotopes of samarium and europium'' by Nowicki et al. reported the observation of $^{138}$Sm \cite{1982Now01}. An enriched $^{112}$Sn target was bombarded with a 190 MeV $^{32}$S beam from the Dubna U-300 cyclotron. The reaction products were identified with the on-line BEMS-2 mass separator and by measuring X- and $\gamma$-rays. The observation of $^{138}$Sm is not discussed in detail and the measured half-life of 3.0(3)~min is only listed in a table. This half-life agrees with the presently adopted values of 3.1(2)~min. Nowicki et al. did not consider their observation a new discovery quoting a previous measurement by Westgaard et al. However, these results were only published in a conference proceeding \cite{1973Wes01}.

\subsection*{$^{139}$Sm}\vspace{0.0cm}
van Klinken et al. described the observation of $^{139}$Sm in ``$^{139}$Sm, a new nuclide of relevance for the systematics of the N=77 isomers'' in 1971 \cite{1971van01}. Alpha particles up to 104~MeV from the Karlsruhe isochronous cyclotron bombarded enriched $^{144}$Sm and $^{142}$Nd and $^{139}$Sm was produced in the reactions $^{144}$Sm($\alpha$,$\alpha$5n) and $^{142}$Nd($\alpha$,7n), respectively. Gamma-ray spectra were recorded with Ge(Li) detectors after irradiation. ``Two new activities with half-lives of 2.6$\pm$0.3~min and 9.5$\pm$1.0~s are produced with constant yield ratios from $^{144}$Sm and $^{142}$Nd, but not from $^{141}$Pr and $^{140}$Ce. When produced from $^{142}$Nd, the threshold for these activities is 10$\pm$3~MeV higher than that for the known isotope $^{140}$Sm. Therefore, both activities have to be attributed to A = 139 and Z = 62.'' These half-lives agree with the currently accepted values of 2.57(10)~min and 10.7(6)~s for the ground and isomeric state, respectively. Previous searches for $^{139}$Sm were unsuccessful \cite{1968Ble01,1970Dro01}.

\subsection*{$^{140,141}$Sm}\vspace{0.0cm}
$^{140}$Sm and $^{141}$Sm were discovered in 1967 by Herrmann et al. as reported in the paper ``Neue Isotope $^{141}$Sm und $^{141}$Sm'' \cite{1967Her01}. The Dubna synchrocyclotron was used to bombard metallic erbium with 660~MeV protons. Resulting activities were measured with MST-17 end-window counters following chemical separation. ``The new isotopes of samarium $^{141}$Sm and $^{140}$Sm with the half-life-times of (22.5$\pm$1.4) minutes and (13.7$\pm$0.8) minutes respectively have been obtained by bombarding metallic erbium with 660 MeV protons.'' These half-lives agree with the currently adopted values of 14.82(12)~min and 22.6(2)~min for $^{140}$Sm and $^{141}$Sm, respectively. Previous half-life measurement for $^{141}$Sm of 17.5$-$22~d \cite{1956Pav01} and 40$-$70~min \cite{1959Lav01} were evidently incorrect. A specific search for the former half-life was unsuccessful \cite{1966Lad01}.

\subsection*{$^{142}$Sm}\vspace{0.0cm}
In 1959 Gratot et al. reported the observation of $^{142}$Sm in ``\'Etude de quelques isotopes tr\`es d\'eficients en neutrons du prom\'eth\'eum et du samarium'' \cite{1959Gra01}. Alpha-particles were accelerated with the Saclay cyclotron to 44 MeV and bombarded an enriched $^{142}$Nd target. Beta-decay curves were measured with a Geiger counter. ``En conclusion, les r\'esultats exp\'erimentaux montrent clairement que l'activit\'e de 8.6 minutes provient de la r\'eaction $^{142}$Nd($\alpha$,3n)$^{143}$Sm et que l'activit\'e de 73 minutes provient de la r\'eaction $^{142}$Nd($\alpha$, 4n)$^{142}$Sm.'' [In conclusion, the experimental results clearly show that the 8.6 minute activity is due to the reaction $^{142}$Nd($\alpha$,3n)$^{143}$Sm and that the 73 minute activity is due to the reaction $^{142}$Nd($\alpha$,4n)$^{142}$Sm.] This half-life for $^{142}$Sm agrees with the presently adopted value of 72.49(5)~min. In the 1958 Table of Isotopes a half-life of 72~min was tentatively assigned to $^{142}$Sm based on a private communication \cite{1958Str01}.

\subsection*{$^{143}$Sm}\vspace{0.0cm}
Silva and Goldemberg observed $^{143}$Sm in 1956 as described in ``Photodisintegration of samarium'' \cite{1956Sil01}. Sm$_{2}$O$_{3}$ was irradiated with x-rays from the Sao Paulo 22 MeV betatron. Decay curves, absorption spectra and excitation functions were measured. ``The only two isotopes of Samarium that by neutron emission could result in a nucleus decaying in 9.03 minutes are $^{147}$Sm (15\%) and $^{144}$Sm (3.1\%), the other isotopes giving rise by neutron emission to other well known activities. The threshold for the reaction $^{147}$Sm($\gamma$,n)$^{146}$Sm and $^{144}$Sm($\gamma$,n)$^{143}$Sm computed from the semi-empirical mass formula are 7.08 and 9.31 MeV. Our result of 9.6 MeV supports the previous assignment of $^{144}$Sm as the isotope of Samarium involved in the reaction investigated in this paper.'' The observed half-life of 9.03~min is in agreement with the currently accepted value of 8.75(8)~min. The previous assignment mentioned in the quote refers to two papers by Butement where a half-life of 8~min was assigned to either $^{143}$Sm or $^{146}$Sm \cite{1950But01,1951But01}. In the second paper \cite{1951But01} Butement states that the observation is consistent with an assignment to $^{143}$Sm, however, in the summary table both isotopes are listed as possibilities. Only two weeks later the assignment of a 8.3~min half-life to $^{143}$Sm was independently reported by Mirnik and Aten \cite{1956Mir01}.

\subsection*{$^{144}$Sm}\vspace{0.0cm}
Aston discovered $^{144}$Sm in 1934 in the paper ``The isotopic constitution and atomic weights of the rare earth elements'' \cite{1934Ast03}. Bromide made from oxalate was used as the source for the Cavendish mass spectrograph. ``Samarium is the most complex of the rare earths. The results obtained with the first setting showed five main isotopes and more intense spectra obtained later revealed two more. Photometry gave the following figures: Mass numbers (Percentage abundance) 144 (3), 147 (17), 148 (14), 149 (15), 150(5), 154 (20), corresponding to a mean mass number 150.2$\pm$0.2''

\subsection*{$^{145}$Sm}\vspace{0.0cm}
$^{145}$Sm was discovered in 1947 by Inghram et al. as reported in the paper ``Activities induced by pile neutron bombardment of samarium'' \cite{1947Ing02}. A Sm$_{2}$O$_{3}$ sample was irradiated with neutrons at the Hanford Pile. $^{145}$Sm was subsequently identified by mass spectroscopy. ``The Sm$^{145}$ was formed by ($n$,$\gamma$) reaction on Sm$^{144}$. It probably decays by K-capture or positron emission to 61$^{145}$. Since the ratio of blackening at 145 and 161 positrons is that characteristic of samarium, the half-life of the 61$^{145}$ can not be the same order of magnitude as the half-life of Sm$^{145}$. The half-lives of Sm$^{145}$ and Gd$^{153}$ were shown to be greater than 72 days by comparison with the decay of 72-day Tb$^{160}$.'' This half-life estimate is consistent with the presently accepted value of 340(3)~d.

\subsection*{$^{146}$Sm}\vspace{0.0cm}
``Alpha activity of Sm$^{146}$ as detected with nuclear emulsions'' was published in 1953 by Dunlavey et al. reporting the observation of $^{146}$Sm \cite{1953Dun01}. A neodymium metal target was bombarded with 40 MeV $^4$He from the Berkeley 60-in. cyclotron. Alpha-particle tracks in nuclear photographic emulsions were examined with a microscope. ``An approximation of the total Sm$^{146}$ produced was then made through yield comparisons by calculating the amounts of both Sm$^{153}$ and Sm$^{145}$ initially formed and by estimating the ratio of the amount of Sm$^{146}$ formed to each of these. Correlation with the observed rate of 2.55 MeV alpha-particle emission gives a half-life approximation of 5$\times$10$^{7}$ years for Sm$^{146}$.'' This half-life is within a factor of two of the currently accepted value of 103(5) My.

\subsection*{$^{147-149}$Sm}\vspace{0.0cm}
In the 1933 paper ``Constitution of neodymium, samarium, europium, gadolinium and terbium'' Aston reported the first observation of $^{147}$Sm, $^{148}$Sm, and $^{149}$Sm \cite{1933Ast01}. Rare earth elements were measured with the Cavendish mass spectrograph. ``Samarium (62) gives a strong pair 152, 154 and a triplet 147, 148, 149.''

\subsection*{$^{150}$Sm}\vspace{0.0cm}
Aston discovered $^{144}$Sm in 1934 in the paper ``The isotopic constitution and atomic weights of the rare earth elements'' \cite{1934Ast03}. Bromide made from oxalate was used as the source for the Cavendish mass spectrograph. ``Samarium is the most complex of the rare earths. The results obtained with the first setting showed five main isotopes and more intense spectra obtained later revealed two more. Photometry gave the following figures: Mass numbers (Percentage abundance) 144 (3), 147 (17), 148 (14), 149 (15), 150(5), 154 (20), corresponding to a mean mass number 150.2$\pm$0.2''

\subsection*{$^{151}$Sm}\vspace{0.0cm}
$^{151}$Sm was discovered in 1947 by Inghram et al. as reported in the paper ``Activities induced by pile neutron bombardment of samarium'' \cite{1947Ing02}. A Sm$_{2}$O$_{3}$ sample was irradiated with neutrons at the Hanford Pile. $^{151}$Sm was subsequently identified by mass spectroscopy. ``The Sm$^{151}$ has previously been observed only in fission. It decays with a half-life of roughly 20 years to Eu$^{151}$.'' The currently accepted value is 90(8)~y. The previous observation mentioned in the quote refers to the 1946 Plutonium Project report which tentatively assigned $^{151}$Sm based on internal reports \cite{1946TPP01}. A previous assignment of a 21-min. half-life to $^{151}$Sm \cite{1938Poo02} was evidently incorrect.

\subsection*{$^{152}$Sm}\vspace{0.0cm}
In the 1933 paper ``Constitution of neodymium, samarium, europium, gadolinium and terbium'' Aston reported the first observation of $^{147}$Sm, $^{148}$Sm, and $^{149}$Sm \cite{1933Ast01}. Rare earth elements were measured with the Cavendish mass spectrograph. ``Samarium (62) gives a strong pair 152, 154 and a triplet 147, 148, 149.''

\subsection*{$^{153}$Sm}\vspace{0.0cm}
The first detection of $^{153}$Sm was reported in 1938 by Pool and Quill in ``Radioactivity induced in the rare earth elements by fast neutrons'' \cite{1938Poo02}. Fast and slow neutrons were produced with 6.3 MeV deuterons from the University of Michigan cyclotron. Decay curves were measured with a Wulf string electrometer. ``These data strongly suggest that the 21-min. period should be assigned to Sm$^{151}$ and the 46-hr. period to Sm$^{153}$.'' The observed half-life of 46~h agrees with the currently accepted value of 46.284(4)~h.

\subsection*{$^{154}$Sm}\vspace{0.0cm}
In the 1933 paper ``Constitution of neodymium, samarium, europium, gadolinium and terbium'' Aston reported the first observation of $^{147}$Sm, $^{148}$Sm, and $^{149}$Sm \cite{1933Ast01}. Rare earth elements were measured with the Cavendish mass spectrograph. ``Samarium (62) gives a strong pair 152, 154 and a triplet 147, 148, 149.''

\subsection*{$^{155}$Sm}\vspace{0.0cm}
The observation of $^{155}$Sm by Winsberg was published as part of the Plutonium Project in 1951: ``Study of 25m Sm$^{(155)}$ in fission'' \cite{1951Win02}. U$^{235}$ and Pu$^{239}$ were irradiated with neutrons at the Los Alamos Homogenous Pile. Decay curves and absorption spectra were measured following chemical separation. ``In order to fit on the fission-yield$-$mass curves for U$^{235}$ and Pu$^{239}$, this samarium species must have a mass of 155. Since 2y Eu$^{155}$ is already known, this is strong evidence for the existence of the following chain: 25min Sm$^{155} \rightarrow$ 2y Eu$^{155} \rightarrow$ stable Gd$^{155}$.'' This half-life agrees with the presently adopted value of 22.3(2)~min. Earlier a 21~min half-life was assigned to $^{151}$Sm \cite{1938Poo02} and it was also published only as a conference abstract without a mass assignment \cite{1942Kur01}.

\subsection*{$^{156}$Sm}\vspace{0.0cm}
$^{156}$Sm was reported as part of the Plutonium Project by Winsberg in 1951 in the article ``Study of the fission chain 10h Sm$^{(156)}$ 15.4d Eu$^{(156)}$'' \cite{1951Win03}. Uranyl nitrate was irradiated with neutrons at the Argonne Heavy-water Pile. Decay curves and absorption spectra were recorded following chemical separation. A mass assignment of 160 was made based on the smooth fission-yield$-$mass curve. ``The fission yield of the $\sim$10h Sm, as determined from the decay curves (about 0.012 per cent), is approximately the same as the fission yield of the 15.4d Eu, thus establishing the following chain relationship: 10h Sm$^{(156)}\rightarrow$15.4d Eu$^{(156)}\rightarrow$ stable Gd$^{(156)}$.'' This half-life agrees with the currently accepted value of 9.4(2)~h.

\subsection*{$^{157}$Sm}\vspace{0.0cm}
D'Auria et al. identified $^{157}$Sm in the 1973 paper ``The decay of $^{160}$Eu'' \cite{1973DAu01}. Natural gadolinium metal chips and enriched $^{160}$Gd samples were irradiated with 14.8 MeV neutrons from a TNC neutron generator and $^{157}$Sm was produced in (n,$\alpha$) reactions. X- and $\gamma$-rays were measured with Ge(Li) spectrometers.  ``A second component of 8$\pm$1 min is observed in these studies associated with several $\gamma$ rays. The most likely assignment appears to be the decay of $^{157}$Sm, but this could not be determined unambiguously.'' This half-life agrees with the presently accepted value of 8.03(7)~min. Previously reported half-lives of 0.5(1)~min \cite{1960Wil02} and 83(2)~s \cite{1972Mor01} were evidently incorrect. Less than 4 months later Kaffrell independently reported a 8.0(5)~min half-life for $^{157}$Sm \cite{1973Kaf01}.

\subsection*{$^{158}$Sm}\vspace{0.0cm}
$^{158}$Sm was discovered by Wilhelmy et al. in 1970 in ``Ground-state bands in neutron-rich even Te, Xe, Ba, Ce, Nd, and Sm isotopes produced in the fission of $^{252}$Cf'' \cite{1970Wil01}. The isotope was observed in the spontaneous fission of a $^{252}$Cf source. The identification was based on coincidence measurements of both fission fragments and K-x-rays. Gamma-rays for several isotopes were measured and the results were only displayed in a table. The first 4 transitions up to the decay of the 8$^+$ state were measured for $^{158}$Sm.

\subsection*{$^{159,160}$Sm}\vspace{0.0cm}
$^{159}$Sm and $^{160}$Sm were discovered in 1986 by Mach et al. with the results published in their paper titled ``Identification of four new neutron rare-earth isotopes'' \cite{1986Mac01}. $^{159}$Sm and $^{160}$Sm were produced in thermal neutron fission of $^{235}$U at Brookhaven National Laboratory. X-rays and $\gamma$-rays were measured at the on-line mass separator TRISTAN. ``Four new neutron-rich, fission-product nuclei have been identified at the on-line mass separator TRISTAN at Brookhaven National Laboratory. Their half-lives have been measured to be, for $^{156}$Pm, T$_{1/2}$ = 28.2$\pm$l.l~sec; for $^{159}$Sm, T$_{1/2}$ = 15$\pm$2~sec; for $^{160}$Sm, T$_{1/2}$ = 8.7$\pm$1.4~sec; and for $^{161}$Eu, T$_{1/2}$ = 27$\pm$3~sec.'' The measured half-lives of 15(2)~s ($^{159}$Sm) and 8.7(14)~s ($^{160}$Sm) are consistent with the currently adopted values of 11.37(15)~s and 9.6(3)~s, respectively.

\subsection*{$^{161}$Sm}\vspace{0.0cm}
$^{161}$Sm was observed in 1998 by Ichikawa et al. as described in ``Identification of $^{161}$Sm and $^{165}$Gd'' \cite{1998Ich01}. $^{238}$U targets were bombarded with 20 MeV protons from the JAERI tandem accelerator facility. $^{161}$Sm was separated with a gas-jet coupled thermal ion source system in the JAERI-ISOL. Beta- and gamma-rays were measured with plastic scintillator and Ge detectors, respectively. ``The new isotope $^{161}$Sm was identified by observing the K x rays of Eu in the A=177 fraction, as shown in the inset of [the figure]. The $\gamma$ ray of 263.7 keV coincident with the Eu K x rays is associated with the $\beta^-$ decay of $^{161}$Sm. The intensities of the Eu K$_{\alpha}$ X and 263.7 keV $\gamma$ rays decayed with half-lives of 5.0$\pm$0.6 s, and 4.4$\pm$0.8 s, respectively as shown in [the figure].'' The quoted half-life of 4.8(8)~s is corresponds to the currently accepted value.

\subsection*{$^{162}$Sm}\vspace{0.0cm}
Ichikawa et al. reported the first observation of $^{162}$Sm in ``$\beta$-decay half-lives of new neutron-rich rare-earth isotopes $^{159}$Pm, $^{162}$Sm, and $^{166}$Gd'' in 2005 \cite{2005Ich01}. $^{238}$U targets were bombarded with 15.5 MeV protons from the JAERI tandem accelerator facility. $^{162}$Sm was separated with a gas-jet coupled thermal ion source system in the JAERI-ISOL. Beta- and X/gamma-rays were measured with a sandwich-type plastic scintillator and two Ge detectors, respectively. ``In addition to these $\gamma$ transitions, previously unobserved $\gamma$ lines having a half-life of $\sim$2.5 s and energies of 36.0, 736.6, and 741.1 keV are also seen in the spectrum. None of these lines are associated with the decay of the contaminant isotopes already identified. Thus, we assign these new transitions to the $\beta^-$ decay of $^{162}$Sm.'' The quoted half-life of 2.4(5)~s corresponds to the presently accepted value.

\section{Discovery of $^{130-165}$Eu}

Thirty-four europium isotopes from A = 130--165 have been discovered so far; these include 2 stable, 19 neutron-deficient and 13 neutron-rich isotopes. According to the HFB-14 model \cite{2007Gor01}, on the neutron-rich side the bound isotopes should reach at least up to $^{210}$Eu. The proton drip has been crossed with the observation of proton emission from $^{130}$Eu and $^{131}$Eu. $^{132}$Eu and $^{133}$Eu  still have to be observed. In addition, four more ($^{126-129}$Eu) could still have half-lives longer than 10$^{-9}$~ns \cite{2004Tho01}. Thus, about 51 isotopes have yet to be discovered corresponding to 60\% of all possible europium isotopes.

\begin{figure}
	\centering
	\includegraphics[scale=.7]{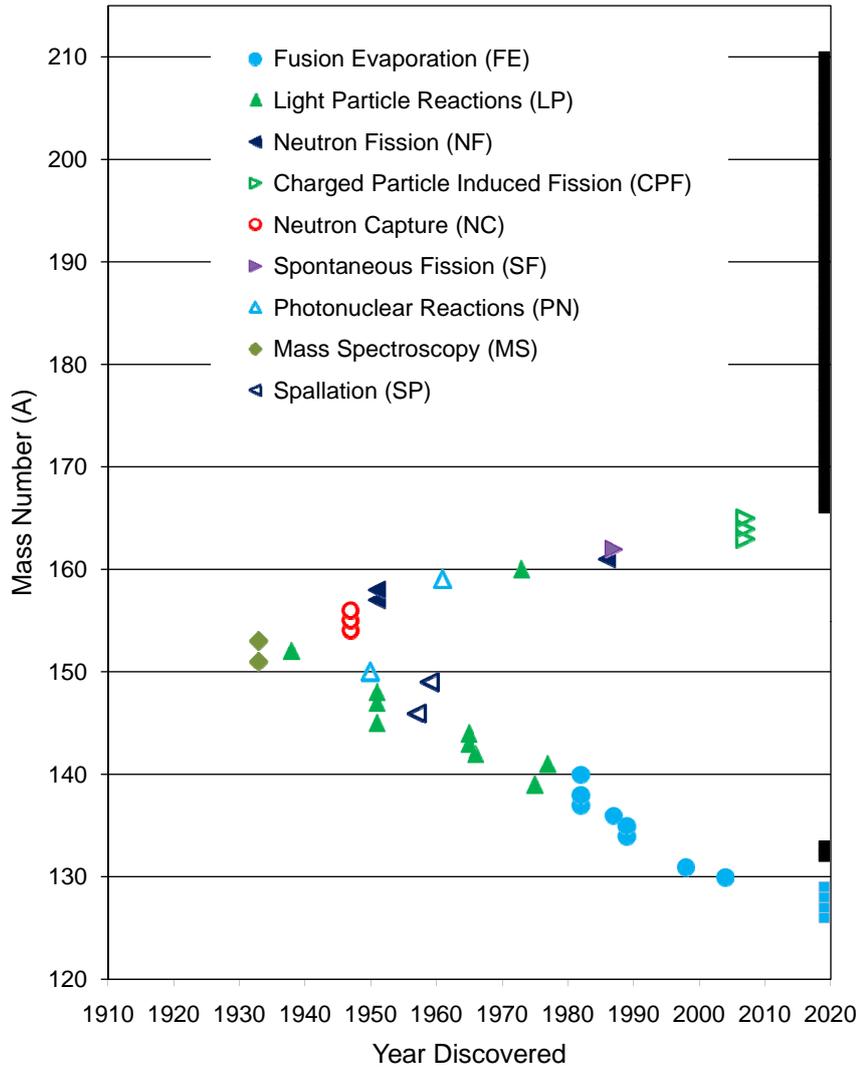}
	\caption{Europium isotopes as a function of time when they were discovered. The different production methods are indicated. The solid black squares on the right hand side of the plot are isotopes predicted to be bound by the HFB-14 model. On the proton-rich side the light blue squares correspond to unbound isotopes predicted to have lifetimes larger than $\sim 10^{-9}$~s.}
\label{f:year-eu}
\end{figure}

Figure \ref{f:year-eu} summarizes the year of discovery for all europium isotopes identified by the method of discovery. The range of isotopes predicted to exist is indicated on the right side of the figure. The radioactive europium isotopes were produced using fusion evaporation reactions (FE), light-particle reactions (LP), neutron induced fission (NF), charged-particle induced fission (CPF), spontaneous fission (SF), neutron capture (NC), photo-nuclear reactions (PN), and spallation (SP). The stable isotopes were identified using mass spectroscopy (MS). Light particles also include neutrons produced by accelerators. In the following, the discovery of each europium isotope is discussed in detail.

\subsection*{$^{130}$Eu}\vspace{0.0cm}
The 2004 paper ``Proton decay of the highly deformed odd-odd nucleus $^{130}$Eu'' by Davids et al. reported the discovery of $^{130}$Eu \cite{2004Dav01}.  A $^{58}$Ni target was bombarded with a 432 MeV $^{78}$Kr beam from the Argonne ATLAS accelerator system and $^{130}$Eu was formed in the fusion evaporation reaction $^{58}$Ni($^{78}$Kr,p5n). Reaction products were separated with the Fragment Mass Analyzer and implanted in a double-sided silicon strip detector where subsequent protons were recorded. ``A peak is clearly visible around $\sim$1 MeV containing six events, with no background. The events are too short lived to be $\beta$ related, and are too low in energy to be from an $\alpha$ decay. The peak is assigned to the new proton emitter $^{130}$Eu.'' The observed half-life of 900$^{+490}_{-290}$ $\mu$s is in agreement with the currently accepted value of 1.1(5)~ms. Preliminary results had been published in a conference proceeding \cite{2002Mah01}.

\subsection*{$^{131}$Eu}\vspace{0.0cm}
Davids et al. observed $^{131}$Eu in 1998 and published their results in ``Proton radioactivity from highly deformed nuclei'' \cite{1998Dav01}. A 222 MeV $^{40}$Ca beam from the Argonne ATLAS accelerator bombarded a $^{96}$Ru target and $^{131}$Eu was formed in the fusion evaporation reaction $^{96}$Ru($^{40}$Ca,p4n). Reaction products were separated with the Fragment Mass Analyzer and implanted in a double-sided silicon strip detector where subsequent protons were recorded.  ``A peak clearly visible at an energy of 950(8) keV (also calibrated using the $^{147}$Tm ground-state proton decay line) which we assign to the proton decay of $^{131}$Eu, produced with a cross section $\sigma \sim$ 90~nb.'' The observed half-life of 26(6)~ms is consistent with the currently accepted value of 17.8(19) ms.

\subsection*{$^{134,135}$Eu}\vspace{0.0cm}
$^{134}$Eu and $^{135}$Eu were observed in 1989 by Veirinen et al. as reported in ``Decay of neutron deficient Eu, Sm, and Pm isotopes near the proton dripline'' \cite{1989Vie02}. An enriched $^{92}$Mo target was bombarded with $^{46}$Ti beams of 204 and 223 MeV  from the Berkeley SuperHILAC. Reaction products were separated with the OASIS on-line mass separator. Charged particles and $\gamma$-rays were measured at the end of a fast-cycling tape system. ``$^{134}_{63}$Eu$_{71}\rightarrow^{134}_{62}$Sm$_{72}$: Beta-delayed proton emission from $^{134}$Eu was established on the basis of proton-Sm K$_\alpha$ X-ray coincidences. Half-life analyses for the protons were carried out with least-squares and maximum-likelihood methods. The former method gave T$_{1/2}$ = 0.5$\pm$0.2~s and the latter T$_{1/2}$ = 0.4$\pm^{0.3}_{0.1}$~s, resulting in an adopted half-life of 0.5$\pm$0.2~s for the $\beta$-delayed proton decay... $^{135}_{63}$Eu$_{71}\rightarrow^{135}_{62}$Sm$_{72}$: The isotope $^{135}$Eu was observed for the first time and identified by Sm K$_\alpha$ X-rays following its EC decay. A half-life of 1.5$\pm$0.2~s was measured for the Sm K$_\alpha$ X-rays.''. These half-lives correspond to the currently accepted values.

\subsection*{$^{136}$Eu}\vspace{0.0cm}
Kern et al. reported the observation of $^{136}$Eu in the 1987 paper ``Transition through triaxial shapes of the light samarium isotopes and the beta decay of $^{136,138,140}$Eu'' \cite{1987Ker01}. A 220 MeV $^{48}$Ti beam from the Oak Ridge HHIRF tandem accelerator bombarded an enriched $^{92}$Mo target. Reaction products were separated with the UNISOR mass separator and $\gamma$-rays were measured with two Ge detectors. ``The assignment of the yrast cascade to A = 136 is confirmed by the magnetic separator results in which the 536-433-256-keV $\gamma$ ray cascade was obtained. We have measured the half-life of the $\beta$ decay of $^{136}$Eu for the first time; a typical decay curve with half-life of 3.9$\pm$0.5~s is shown in [the figure].''. This half-life agrees with the presently accepted values of the ground state (3.3(3)~s) or an isomeric state (3.8(3)~s).

\subsection*{$^{137,138}$Eu}\vspace{0.0cm}
``Very neutron deficient isotopes of samarium and europium'' by Nowicki et al. reported the discovery of $^{137}$Eu and $^{138}$Eu \cite{1982Now01}. An enriched $^{112}$Sn target was bombarded with a 190 MeV $^{32}$S beam from the Dubna U-300 cyclotron. The reaction products were identified with the on-line BEMS-2 mass separator and by measuring X- and $\gamma$-rays. ``Three isotopes: $^{136}$Sm, $^{137}$Eu, and $^{138}$Eu (with half-lives 40$\pm$5 s, 11$\pm$2 s and 12$\pm$2 s respectively) were observed for the first time''. These half-lives are in agreement with the currently accepted values of 8.4(5)~s and 12.1(6)~s for $^{137}$Eu and $^{138}$Eu, respectively. Previously reported half-lives of 1.5(4)~s and 35(6)~s \cite{1977Bog01} were evidently incorrect.

\subsection*{$^{139}$Eu}\vspace{0.0cm}
$^{139}$Eu was observed by van Klinken and Feenstra in 1975 as reported in ``Shape implications of unhindered 11/2$^-$ $\rightarrow$ 11/2$^-$ $\beta$ decays in the region with N$<$82 and Z$>$50'' \cite{1975van01}. Alpha beams accelerated to 140 MeV by the KVI cyclotron bombarded enriched $^{144}$Sm and $^{142}$Nd, and natural praseodymium targets. Gamma- and beta-rays were measured at the end of a pneumatic transport system. ``Similarly an 11/2$^-$ $\rightarrow$ 11/2$^-$ branch of more than 50\% is proposed for the decay of $^{139}$Eu, an isotope previously known only through its 112-keV transition.'' The 22~s half-life noted in a level scheme is consistent with the currently accepted value of 17.9(6)~s. The previous observation of the 112~keV transition mentioned in the quote and a half-life of 22(3)~s was only published in a conference proceeding \cite{1973Wes01}.

\subsection*{$^{140}$Eu}\vspace{0.0cm}
``Very neutron deficient isotopes of samarium and europium'' by Nowicki et al. reported the observation of $^{140}$Eu \cite{1982Now01}. An enriched $^{112}$Sn target was bombarded with a 190 MeV $^{32}$S beam from the Dubna U-300 cyclotron. The reaction products were identified with the on-line BEMS-2 mass separator and by measuring X- and $\gamma$-rays. The observation of $^{140}$Eu is not discussed in detail and previously measured half-lives of 1.3(2)~s and 20~s were approximately confirmed as listed in a table. The 1.3(2)~s half-life agrees with the presently adopted values of 1.51(2)~s. Nowicki et al. did not consider their observation a new discovery quoting a previous measurement by Westgaard et al. However, these results were only published in a conference proceeding \cite{1973Wes01}. The 20$^{+15}_{-10}$~s half-life previously reported \cite{1972Hab01} was evidently incorrect.

\subsection*{$^{141}$Eu}\vspace{0.0cm}
Deslauriers et al. reported the observation of $^{141}$Eu in the 1977 paper ``The $^{141}$Eu nuclide and its decay properties'' \cite{1977Des01}. Enriched $^{144}$Sm$_{2}$O$_{3}$ targets were bombarded with 35-65 MeV protons from the McGill synchrocyclotron and $^{141}$Eu was produced in (p,4n) reactions. Positrons, x- and $\gamma$-rays were measured with a plastic $\Delta$E$-$E scintillation detector telescope, a Ge x-ray spectrometer and a Ge(Li) detector, respectively. ``The nuclide $^{141}$Eu is identified to have two beta decaying isomers, $^{141g}$Eu and $^{141m}$Eu, whose decay half-lives are measured to be 40.0$\pm$0.7 s and 3.3$\pm$0.3 s, respectively''. The 40.0(7)~s half-life corresponds to the presently adopted value for the ground state and the 3.3(3)~s agrees with the value of 2.7(3)~s for an isomeric state. A previously reported half-life of 37(3)~s was only published in a conference proceeding \cite{1973Wes01}.

\subsection*{$^{142}$Eu}\vspace{0.0cm}
$^{142}$Eu was observed in 1966 by Malan et al. as reported in ``The europium isotopes $^{142}$Eu, $^{143}$Eu, and $^{144}$Eu'' \cite{1966Mal01}. The Karlsruhe cyclotron was used to bombard enriched $^{144}$Sm$_2$O$_3$ targets with 15$-$40 MeV deuterons and $^{142}$Eu was formed in the reaction $^{144}$Sm(d,4n). Gamma-ray spectra were measured with two NaI(Tl) detectors following chemical separation. ``Both for chemical and genetic reasons the one-minute component is therefore identified as $^{142}$Eu. Calculation of the half-lives by means of the IBM computer gave 1.2$\pm$0.2 min for $^{142}$Eu and 72.49$\pm$0.05 min for $^{142}$Sm.'' This half-life agrees with the currently accepted value of 1.223(8)~min.

\subsection*{$^{143}$Eu}\vspace{0.0cm}
In 1965 Kotajma et al.reported the discovery of $^{143}$Eu in ``A new nuclide europium 143'' \cite{1965Kot01}. Samarium oxide targets were irradiated with 26 MeV deuterons from the Amsterdam synchrocyclotron and $^{143}$Eu was formed in the reaction $^{144}$Sm(d,3n). The resulting activities were measured with a scintillation spectrometer following chemical separation. ``A new nucleide europium 143, produced by the (d,3n) reaction on $^{144}$Sm, was investigated. The half-life of the nucleide was found to be T$_{1/2}$=2.3$\pm$0.2~min and the end-point energy of the beta ray was found to be E$_{\beta}$=4.0$\pm$0.2~MeV''. This half-life is consistent with the currently accepted value of 2.59(2)~min.

\subsection*{$^{144}$Eu}\vspace{0.0cm}
``Europium-144'' was published in 1965 by Messlinger et al. reporting the discovery of $^{144}$Eu \cite{1965Mes01}. An enriched $^{144}$Sm target was bombarded with protons from the Heidelberg Tandem Accelerator. Decay curves and $\gamma$-spectra were measured with a NaI(Tl) scintillation spectrometer. ``In the course of measurements of (p,n) cross sections on N=82 nuclei, we came across a short-live activity which we assign to $^{144}_{63}$Eu$_{81}$. This newly found activity is analogous to the ground state of $^{140}_{59}$Pr$_{81}$ and $^{142}_{61}$Pm$_{81}$.'' The observed half-life of 10.5(3)~s agrees with the currently accepted value of 10.2(1)~s. A 18-min half-life listed in the 1958 table of isotopes \cite{1958Str01} was based on a private communication and evidently incorrect \cite{1959Olk01}.

\subsection*{$^{145}$Eu}\vspace{0.0cm}
Hoff et al. reported the observation of $^{145}$Eu in the 1951 paper ``Neutron deficient europium and gadolinium isotopes'' in 1951 \cite{1951Hof01}. Enriched $^{147}$Sm$_2$O$_3$ and gadolinium oxide targets were bombarded with 50 MeV and 150 MeV protons, respectively. $^{145}$Eu was then observed in the $^{147}$Sm(p,3n) reaction or in the $\alpha$ decay of $^{149}$Tb. Decay curves and absorption spectra were measured following chemical separation. ``Another isotope of europium, Eu$^{145}$, was observed as recoil nuclei from the alpha-decay of Tb$^{149}$. The terbium was produced in a 150-MeV proton bombardment of gadolinium oxide and isolated using a cation exchange column separation. The Eu$^{145}$ has also been produced in a 50-MeV proton bombardment of Sm$_{2}^{147}$O$_{3}$''. The observed half-life of 5(1)~d is in agreement with the currently accepted value of 5.93(4)~d.

\subsection*{$^{146}$Eu}\vspace{0.0cm}
$^{146}$Eu was reported by Gorddinskii et al. in the 1957 paper ``Neutron-deficient isotopes of the rare earth elements forming as a result of spallation of Ta under bombardment with 660 MeV protons'' \cite{1957Gor01}. A tantalum target was bombarded with 660 MeV protons from the JINR synchrocyclotron. Resultant activities were measured following chromatographic separation. ``According to our revised data, the $\gamma$-spectrum of ~5 day Eu$^{145}$ consists of a 636 and 745 kev line with an intensity ratio of 1.0:2.3.'' A note added in proof stated ``We are now of the opinion that the Gd and Eu isotopes described in this section actually have mass number 146.'' This half-life agrees with the currently accepted value of 4.61(3)~d. A previously measured half-life of 38(3)~h \cite{1951Hof01} was evidently incorrect.

\subsection*{$^{147,148}$Eu}\vspace{0.0cm}
Hoff et al. reported the observation of $^{147}$Eu and $^{148}$Eu in the 1951 paper ``Neutron deficient europium and gadolinium isotopes'' in 1951 \cite{1951Hof01}. Enriched $^{147}$Sm and $^{148}$Sm targets were bombarded with 8.5 MeV protons producing $^{147}$Eu and $^{148}$Eu, respectively, in (p,n) charge exchange reactions. Decay curves and absorption spectra were measured. The decay of $^{147}$Eu was studied following chemical separation. ``The decay of europium isotopes with mass numbers 147 and 148 was observed after proton bombardments of enriched samarium isotopes. The (p,n) reaction is probably virtually the only nuclear reaction induced by 8.5-MeV protons on samarium.''. The observed half-lives of 24(2)~days ($^{147}$Eu) and 50(2)~d ($^{148}$Eu) are in agreement with the currently accepted values of 24.1(6)~d and 54.5(5)~d, respectively. Previous tentative assignments of 54~d \cite{1950Wil04} and 53~d \cite{1951Mar02} half-lives to $^{147}$Eu were incorrect.

\subsection*{$^{149}$Eu}\vspace{0.0cm}
In the 1959 paper ``Conversion electrons from Eu$^{149}$'' Antoneva et al. described the observation of $^{149}$Eu \cite{1959Ant01}. Tantalum targets were bombarded with 660 MeV protons and conversion electron spectra of the chemically separated spallation products were measured. ``In investigating the gadolinium fraction, we found that after the decay of Gd$^{147}$ and Gd$^{149}$ there remained in the conversion spectrum lines corresponding to transitions of the above mentioned energies. The intensities of these lines fell off with a like period ($\sim$100 days). Apparently, our gadolinium fraction contained the parent of the europium activity under study. It follows, therefore, that this activity cannot be attributed to Eu$^{147}$, in as much as the half-life of Gd$^{148}$ is greater than 35 years and, consequently, if the activity were due to Eu$^{147}$ the intensity of our conversion lines would increase with time instead of decreasing. Thus it maybe safely be inferred that the observed conversion electrons are emitted in the Eu$^{149}\rightarrow$Sm$^{149}$ decay process.'' The observed half-life of $\sim$120~d is close to the currently accepted value of 93.1(4)~d. Previous tentative assignments of a 14~d half-life \cite{1950Wil04,1951Mar02} to $^{149}$Eu were evidently incorrect.

\subsection*{$^{150}$Eu}\vspace{0.0cm}
$^{150}$Eu was observed by Butement in 1950 as reported in ``New radioactive isotopes produced by nuclear photo-disintegration'' \cite{1950But01}. $^{150}$Eu was produced through irradiation of europium oxide by 23~MeV x-rays from a synchrotron in the photonuclear reaction $^{151}$Eu($\gamma$,n) and chemically separated from other resultant isotopes \cite{1950But03}. In the original paper \cite{1950But01} a probable assignment was only given in a table. More details were reported in the subsequent publication \cite{1950But03}: ``The activity showed a nearly logarithmic decay with an apparent half-life of 12 hours. Such a decay, resulting from two slightly different half-lives, is
difficult to resolve accurately unless one component is known. By assuming one component to be 9.2-hour $^{152}$Eu which must have been formed, the best value for the half-life of the other component was 15 hours, and its yield was nearly equal to that of the $^{152}$Eu. The 15-hour activity must almost certainly be $^{150}$Eu, and since the abundances of $^{151}$Eu and $^{153}$Eu are nearly equal, the relative yield indicates a high counting efficiency, so that decay is probably by positron emission.'' This half-life agrees with the presently adopted value of 12.8(1)~h. Previously Pool and Quill had tentatively assigned a 27-h half-life to $^{150}$Eu \cite{1938Poo02} which differs significantly from the correct value.

\subsection*{$^{151}$Eu}\vspace{0.0cm}
In the 1933 paper ``Constitution of neodymium, samarium, europium, gadolinium and terbium'' Aston reported the first observation of $^{151}$Eu \cite{1933Ast01}. Rare earth elements were measured with the Cavendish mass spectrograph. ``Europium (63), as expected from its chemical atomic weight (152.0), contains the two odd mass numbers 151, 153 in roughly equal abundance.''

\subsection*{$^{152}$Eu}\vspace{0.0cm}
The first detection of $^{152}$Eu was reported in 1938 by Pool and Quill in ``Radioactivity induced in the rare earth elements by fast neutrons'' \cite{1938Poo02}. Fast and slow neutrons were produced with 6.3 MeV deuterons from the University of Michigan cyclotron. Decay curves were measured with a Wulf string electrometer. ``The 9.2-hr. period is assigned to Eu$^{152}$ instead of to Eu$^{154}$ since this period is strongly produced by fast neutron bombardment.'' This half-life is in agreement with the currently accepted value of 9.3116(13)~h. Previously a 9.2(1)~h half-life was assigned to either $^{152}$Eu or $^{154}$Eu \cite{1935Mar01}.

\subsection*{$^{153}$Eu}\vspace{0.0cm}
In the 1933 paper ``Constitution of neodymium, samarium, europium, gadolinium and terbium'' Aston reported the first observation of $^{153}$Eu \cite{1933Ast01}. Rare earth elements were measured with the Cavendish mass spectrograph. ``Europium (63), as expected from its chemical atomic weight (152.0), contains the two odd mass numbers 151, 153 in roughly equal abundance.''

\subsection*{$^{154}$Eu}\vspace{0.0cm}
Inghram and Hayden reported the observation of $^{154}$Eu in the 1947 paper ``Artificial activities produced in europium and holmium by slow neutron bombardment'' \cite{1947Ing03}. A nitric acid solution of Eu$_{2}$0$_{3}$ was irradiated with slow neutrons to produce $^{154}$Eu. ``The irradiated sample was allowed to stand for two weeks in order that the 9.2-hour activity might decay. An aliquot of this sample was then run in the mass spectrograph, and a transfer plate was made. Development of the transfer plate showed two lines at masses 152 and 154... Thus europium must have two long lived activities, one of mass 152 and one of mass 154, as well as the established 9.2 hour at mass 152.''

\subsection*{$^{155,156}$Eu}\vspace{0.0cm}
$^{155}$Eu and $^{156}$Eu were discovered in 1947 by Inghram et al. as reported in the paper ``Activities induced by pile neutron bombardment of samarium'' \cite{1947Ing02}. A Sm$_{2}$O$_{3}$ sample was irradiated with neutrons at the Hanford Pile. $^{155}$Eu and $^{156}$Eu were subsequently identified by mass spectroscopy. ``The 2-3 year Eu$^{155}$ was formed by the reaction Sm$^{154}$ $\stackrel{ (n,\gamma)}{\longrightarrow}$ Sm$^{155}$ $\stackrel{\beta}{\longrightarrow}T_{1/2}$=21 min Eu$^{155}$. It has previously only been observed in fission. The 15.4-day Eu$^{156}$ had also been observed only in fission. Here it was formed by (n,$\gamma$) reaction on the Eu$^{155}$.'' These half-lives are close to the currently accepted values of 4.7611(13)~y and 15.19(80)~d for $^{155}$Eu and $^{156}$Eu, respectively. The previous observation mentioned in the quote refers to the 1946 Plutonium Project report which assigned $^{155}$Eu and $^{156}$Eu based on internal reports \cite{1946TPP01}.

\subsection*{$^{157,158}$Eu}\vspace{0.0cm}
The observation of $^{157}$Eu and $^{158}$Eu by Winsberg was published as part of the Plutonium Project in 1951: ``Study of 60m Eu$^{(158)}$ and 15.4h Eu$^{(157)}$ activities in fission'' \cite{1951Win01}. Uranyl nitrate was irradiated with neutrons in the thimble of the Argonne Heavy-water Pile. Decay curves and absorption spectra were measured with a thin-windowed Geiger Mueller tube following chemical separation. ``The tentative mass assignments of 158 for the 60m Eu and of 157 for the 15.4h Eu was suggested by the fission-yield$-$mass curve of U$^{235}$. Neither isotope has been mentioned in the published literature.'' These half-lives are close to the presently adopted values of 15.18(3)~h and 45.9(2)~min for $^{157}$Eu and $^{158}$Eu, respectively.

\subsection*{$^{159}$Eu}\vspace{0.0cm}
Kuroyanagi et al. observed $^{159}$Eu in 1961 as reported in ``New activities in rare earth region produced by the ($\gamma$,p) reactions'' \cite{1961Kur01}. Pure oxide powder was irradiated with $\gamma$-rays at the Tohoku 25 MeV betatron. Decay curves were measured with a beta ray analyser or an end-window G-M counter and $\beta$-ray spectra were recorded with a plastic scintillator. ``There has been no detailed investigation about the nucleide Eu$^{159}$ since Butement first gave its half-life as about 20 min in 1951... After a long time measurement, a long component activity observed in the decay curve shown in [the figure] was proved to be due to the l8-h activity from Gd$^{159}$ and the 15.4-h activity from Eu$^{157}$. Subtracting these long components, we observed an activity of about 19 min.'' The quoted half-life of 19(1)~min agrees with the currently accepted value of 18.1(1)~m. The previous measurement mentioned in the quote was assigned by Butement to either $^{159}$Eu or an isomeric state of stable gadolinium \cite{1951But01}.

\subsection*{$^{160}$Eu}\vspace{0.0cm}
D'Auria et al. identified $^{160}$Eu in the 1973 paper ``The decay of $^{160}$Eu'' \cite{1973DAu01}. Natural gadolinium metal chips and enriched $^{160}$Gd samples were irradiated with 14.8 MeV neutrons from a TNC neutron generator and $^{160}$Eu was produced in (n,p) charge-exchange reactions. X- and $\gamma$-rays were measured with Ge(Li) spectrometers. ``The decay of $^{160}$Eu was studied using fast neutrons on enriched samples of gadolinium. A half-life of 50$\pm$10~s, associated with gamma rays definitely arising from transitions in $^{160}$Gd, is assigned to the decay of $^{160}$Eu.'' This half-life agrees with the presently accepted value of 38(4)~s. A previously reported half-life of $\sim$2.5~min \cite{1961Tak01} was evidently incorrect.

\subsection*{$^{161}$Eu}\vspace{0.0cm}
$^{161}$Eu was discovered in 1986 by Mach et al. with the results published in their paper titled ``Identification of four new neutron rare-earth isotopes'' \cite{1986Mac01}. $^{161}$Eu was produced in thermal neutron fission of $^{235}$U at Brookhaven National Laboratory. X-rays and $\gamma$-rays were measured at the on-line mass separator TRISTAN. `` Five transitions of energy 71.9$\pm$0.2, 91.9$\pm$0.2, 163.7$\pm$0.2, 293.9$\pm$0.3, and 314.3$\pm$0.3 keV were assigned to the decay of $^{161}$Eu on the basis of $\gamma$-$\gamma$ and x-$\gamma$ coincidences and the measured half-lives. Four of them fit immediately into the level scheme of $^{161}$Gd as revealed in the $^{160}$Gd(d,p) reaction. The T$_{1/2}$ value is the average of the measurements for the 71.9-, 91.9-, and 163.7-keV transitions.'' The measured half-lives of 27(3)~s agrees with the currently adopted values of 26(3)~s.

\subsection*{$^{162}$Eu}\vspace{0.0cm}
In 1987, Greenwood et al. identified $^{162}$Eu in the paper entitled ``Identification of new neutron-rich rare-earth isotopes produced in $^{252}$Cf Fission'' \cite{1987Gre01}. Spontaneous fission fragments from a $^{252}$Cf source were measured with the isotope separation on line (ISOL) system at the Idaho National Engineering Laboratory. $^{162}$Eu was identified by mass separation and the measurement of K x-rays. ``Identification of the $^{162}$Eu decay was accomplished in two separate experiments, with collection-counting cycle times of 24 and 28~s. For each experiment, the resulting $\gamma$-ray spectra were observed to contain a strong contamination from $^{146}$La decay, resulting from LaO molecular beam formation in the ion source. Since the $^{146}$La component was dominant, and since it had a half-life value comparable to that of $^{162}$Eu, only the decay of the Gd K x rays could be used to characterize the $^{162}$Eu decay.'' The observed half-life of 10.6(10)~s corresponds to the currently accepted value. Mach et al. had previously reported weak evidence for a $\sim$6 s half-life and did not claim discovery \cite{1986Mac01}.

\subsection*{$^{163-165}$Eu}\vspace{0.0cm}
Hayashi et al. observed $^{163}$Eu, $^{164}$Eu, and $^{165}$Eu as reported in the 2007 paper ``Q$_\beta$ measurements of $^{158,159}$Pm, $^{159,161}$Sm, $^{160-165}$Eu, $^{163}$Gd, and $^{166}$Tb using a total absorption BGO detector'' \cite{2007Hay01}. A uranium carbide target was bombarded by 32 MeV protons from the Tokai tandem accelerator. $^{163}$Eu, $^{164}$Eu, and $^{165}$Eu were formed in the reaction $^{238}$U(p,f) and identified with the Tokai-ISOL on-line mass separator. Beta-decay energies were measured with two BGO scintillation detectors. ``Isotopes of $^{163,164,165}$Eu were recently identified by our group using this ion source, and the Q$_{\beta}$ values of $^{160-165}$Eu and $^{163}$Gd were measured for the first time.'' The earlier publication mentioned in the quote was a conference proceeding \cite{2006Sat01}.

\section{Discovery of $^{135-166}$Gd}

Thirty-one gadolinium isotopes from A = 135--166 have been discovered so far; these include 7 stable, 17 neutron-deficient and 7 neutron-rich isotopes.
According to the HFB-14 model \cite{2007Gor01}, $^{209}$Gd should be the last odd-even particle stable neutron-rich nucleus while the even-even particle stable neutron-rich nuclei should continue through $^{218}$Gd. At the proton drip line six more particle stable isotopes are predicted ($^{130-134}$Gd, $^{136}$Gd). In addition, five more isotopes ($^{125-129}$Gd) could possibly still have half-lives longer than 10$^{-9}$~ns \cite{2004Tho01}. Thus, about 59 isotopes have yet to be discovered corresponding to 66\% of all possible gadolinium isotopes.

\begin{figure}
	\centering
	\includegraphics[scale=.7]{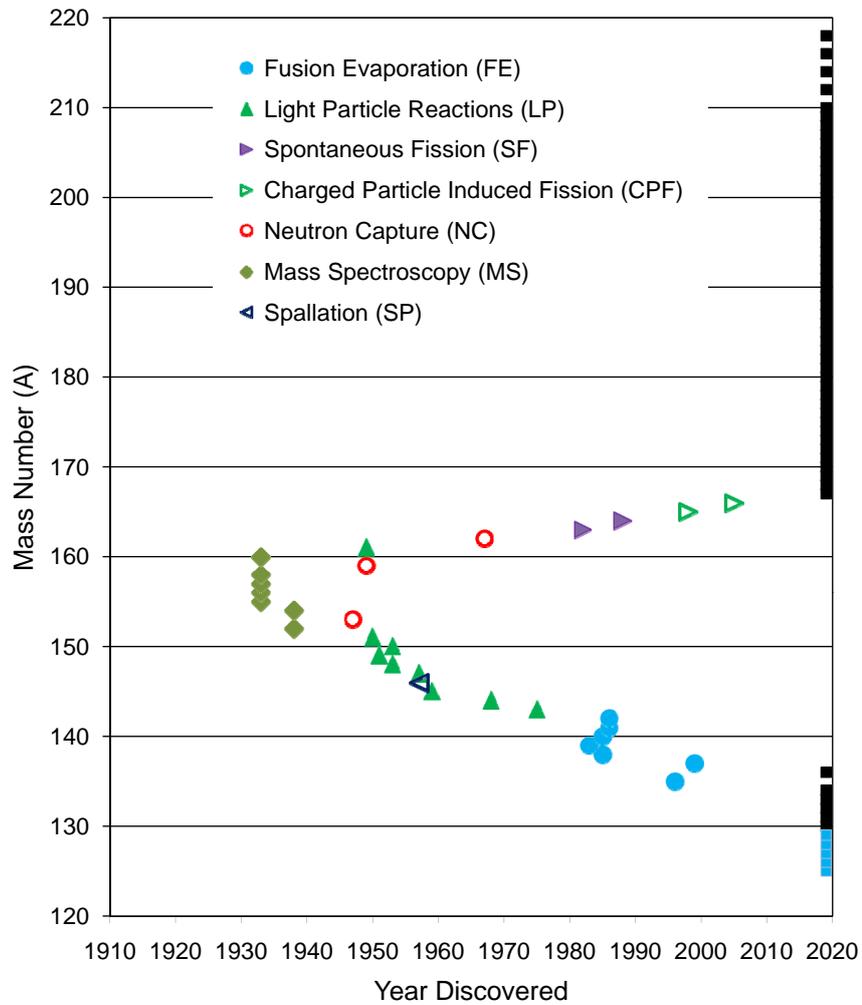}
	\caption{Gadolinium isotopes as a function of time when they were discovered. The different production methods are indicated. The solid black squares on the right hand side of the plot are isotopes predicted to be bound by the HFB-14 model. On the proton-rich side the light blue squares correspond to unbound isotopes predicted to have lifetimes larger than $\sim 10^{-9}$~s.}
\label{f:year-gd}
\end{figure}

Figure \ref{f:year-gd} summarizes the year of discovery for all gadolinium isotopes identified by the method of discovery. The range of isotopes predicted to exist is indicated on the right side of the figure. The radioactive gadolinium isotopes were produced using fusion evaporation reactions (FE), light-particle reactions (LP), charged-particle induced fission (CPF), spontaneous fission (SF), neutron capture (NC), and spallation (SP).  The stable isotope was identified using mass spectroscopy (MS). Light particles also include neutrons produced by accelerators. In the following, the discovery of each gadolinium isotope is discussed in detail.

\subsection*{$^{135}$Gd}\vspace{0.0cm}
Xu et al. published the observation of $^{135}$Gd in the 1996 paper ``New $\beta$-delayed proton precursor $^{135}$Gd'' \cite{1996Xu01}. An enriched $^{106}$Cd target was bombarded with a 171 MeV $^{32}$S beam from the Lanzhou cyclotron and $^{135}$Gd was formed in the reaction $^{106}$Cd($^{32}$S,3$n$). Charged-particle, $\gamma$-rays, and X-rays were measured with a silicon surface barrier, HpGe(GMX) and planar HpGe detector, respectively. ``Unknown $\beta$-delayed proton precursor $^{135}$Gd has been synthesized by the reaction $^{106}$Cd($^{32}$S,3$n$) and identified using a helium-jet recoil fast-tape-transport system with a $\gamma$(X)-p coincidence measurements. Its $\beta$-delayed proton spectrum has been observed. The half-life of $^{135}$Gd decay has been determined to be 1.1$\pm$0.2~s.'' This half-life corresponds to the currently accepted value.

\subsection*{$^{137}$Gd}\vspace{0.0cm}
Xu et al. identified $^{137}$Gd in 1999 as described in ``New $\beta$-delayed proton precursors in the rare-earth region near the proton drip line'' \cite{1999Xu01}. A 220~MeV $^{36}$Ar beam was accelerated with the Lanzhou sector-focused cyclotron and bombarded an enriched $^{106}$Cd target. Proton-$\gamma$ coincidences were measured in combination with a He-jet type transport system. ``From the decay curve of the proton coincident 255-keV $\gamma$ line, the half-life of the $^{137}$Gd decay was extracted to be 2.2$\pm$0.2~s.'' This half-life corresponds to the currently accepted value. A previously measured half-life of 7(3)~s \cite{1983Nit01} was evidently incorrect.

\subsection*{$^{138}$Gd}\vspace{0.0cm}
In 1985 $^{138}$Gd was identified by Lister et al. in ``Deformation of very light rare-earth nuclei'' \cite{1985Lis01}. A $^{50}$Cr beam from the Daresbury Laboratory Van de Graaff accelerator was incident on a $^{94}$Mo target at 220 and 230 MeV. Gamma rays, neutrons and charged particles were detected and new ground-state bands observed. ``This letter reports results on the ground-state bands in the even-even nuclei $^{128}_{58}$Ce$_{68}$, $^{128,130,132}_{60}$Nd$_{68,70,72}$, $^{134,136}_{62}$Sm$_{72,74}$, and $^{138,140}_{64}$Gd$_{74,76}$.''

\subsection*{$^{139}$Gd}\vspace{0.0cm}
Nitschke et al. discovered $^{139}$Gd in the 1983 paper ``New beta-delayed proton emitter in the lanthanide region'' \cite{1983Nit01}. A $^{50}$Cr beam from the Berkeley SuperHILAC bombarded a gas-cooled $^{92}$Mo target. $^{139}$Gd was identified with the on-line isotope separator OASIS. ``The identification of this isotope as $^{139}$Gd is well supported by the (Q$_{EC}$-S$_{P}$) value of 6.50 MeV, a large cross section of 70 mb, and the good agreement between the experimental half-life of 4.9$\pm$1.0~s and the calculated value of 6.4~s.'' This half-life is in agreement with the currently accepted values of 5.7(3)~s for the ground state and 4.8(9)~s for an isomeric state.

\subsection*{$^{140}$Gd}\vspace{0.0cm}
In 1985 $^{140}$Gd was identified by Lister et al. in ``Deformation of very light rare-earth nuclei'' \cite{1985Lis01}. A $^{50}$Cr beam from the Daresbury Laboratory Van de Graaff accelerator was incident on a $^{94}$Mo target at 230 MeV. Gamma rays, neutrons and charged particles were detected and new ground-state bands observed. ``This letter reports results on the ground-state bands in the even-even nuclei $^{128}_{58}$Ce$_{68}$, $^{128,130,132}_{60}$Nd$_{68,70,72}$, $^{134,136}_{62}$Sm$_{72,74}$, and $^{138,140}_{64}$Gd$_{74,76}$.''

\subsection*{$^{141}$Gd}\vspace{0.0cm}
Redon et al. described the first observation of $^{141}$Gd in the 1986 paper ``Exotic neutron-deficient nuclei near N=82'' \cite{1986Red01}. Targets of $^{112}$Sn were bombarded with beams of $^{35}$Cl and $^{32}$S beams from the Grenoble SARA accelerator. The residues were separated with an on-line mass separator and a He-jet system. X-ray and $\gamma$-ray spectra were measured. ``In the present work, a number of $\gamma$-rays with half-life of 22$\pm$3~s in perfect agreement with Takahashi gross theory, are present both in $^{32}$S + $^{112}$Sn and in $^{35}$Cl + $^{112}$Sn reactions but with a minor intensity in the latter one.'' This half-life agrees with the currently accepted value of 24.5(5)~s.

\subsection*{$^{142}$Gd}\vspace{0.0cm}
$^{143}$Gd was identified in 1986 by Lunardi et al. in ``First observation of yrast $\gamma$-rays in the N =78 nuclei $^{142}$Gd and $^{141}$Eu'' \cite{1986Lun01}. Samarium targets were bombarded with 65$-$107~MeV $\alpha$-particle beams from the J\"ulich cyclotron and $^{143}$Gd was produced in the fusion-evaporation reaction $^{144}$Sm($\alpha$,6n). Excitation functions were measured and $\gamma$-ray spectra were recorded with Ge detectors. ``We have therefore reinvestigated $\alpha$-induced (xnyp) excitation functions for the N = 82 targets of Sm and Nd with $\alpha$-particle beams from 65$-$107 MeV from the cyclotron at J\"ulich. From the new data we found that the earlier proposed $^{142}$Gd yrast cascade with the 526 keV 2$^+\rightarrow$0$^+$ transition, must be assigned to the $^{141}$Eu isotonic nucleus formed through the $^{144}$Sm($\alpha$,p6n) reaction, whereas a new yrast cascade based on an intense 515 keV line is assigned to the ($\alpha$,6n) channel populating levels in $^{142}$Gd.'' The incorrect assignment of the previously measured yrast cascade mentioned in the quote was published ten years earlier \cite{1976Mar01}.

\subsection*{$^{143}$Gd}\vspace{0.0cm}
Kosanke et al. observed $^{143}$Gd in the 1975 paper ``An improved and modular helium-jet recoil-transport system for the study of short-lived nuclei'' \cite{1975Kos01}. A samarium target was bombarded with a 70~MeV triton beam from the MSU sector-focused cyclotron and $^{143}$Gd was formed in the reaction $^{144}$Sm($^3$H,4n). Gamma-ray spectra were measured with two Ge(Li) detectors in combination with a helium-jet transport system. ``An example of the use of this coincidence arrangement is given in [the figure], which shows some $\gamma$-ray spectra from the decays of 39-s $^{143g}$Gd and l12-s $^{143m}$Gd.'' These half-lives agree with the presently accepted values of 39(2)~s for the ground state and 110.0(14)~s for the isomeric state. Previously an upper half-life limit of 1 min was reported \cite{1968Kel01}.

\subsection*{$^{144}$Gd}\vspace{0.0cm}
The 1968 paper ``Zerfallseigenschaften der Gadolinium-Isotope $^{145}$Gd und $^{144}$Gd'' by Keller and M\"unzel reported the discovery of $^{144}$Gd. An enriched $^{144}$Sm$_2$O$_3$ target was bombarded with 52~MeV $\alpha$-particles from the Karlsruhe isochronous cyclotron and $^{144}$Gd was formed in the reaction $^{144}$Sm($\alpha$,4n). Beta-decay curves and $\gamma$-ray spectra were measured following chemical separation. ``Die Analyse der $\beta$-Abfallkurven ergab eine Halbwertszeit von 4.5$\pm$0.1~min... Auf Grund der chemischen Abtrennung kann diesem Radionuklid die Ordnungszahl Z = 64 und \"uber die Massentrennung die Nukleonenzahl A = 144 zugeschrieben werden.'' [The analysis of the $\beta$-decay curves resulted in a half-life of 4.55$\pm$0.1~min...  An atomic number of Z = 64 and a nucleon number A = 144 can be assigned to this nuclide based on the chemical separation, and mass separation, respectively.] This half-life agrees with the presently adopted value of 4.47(6)~min.

\subsection*{$^{145}$Gd}\vspace{0.0cm}
Grover published ``Mass assignments and some decay characteristics of Gd$^{145}$, Eu$^{145}$, Gd$^{146}$, and Eu$^{146}$'' in 1959 describing the observation of $^{145}$Gd \cite{1959Gro01}. An enriched $^{144}$Sm$_2$O$_3$ target was bombarded with 40~MeV $\alpha$ particles from the Brookhaven 60-in. cyclotron. Gamma-ray spectra were measured with a NaI(Tl) scintillation crystal following chemical separation. ``The mass number of the 25-minute activity is thus most probably 145. Also, it is quite likely that it is a parent of the previously identified 5-day Eu$^{145}$, and is therefore Gd$^{145}$, its formation being consistent with the reaction Sm$^{144}$($\alpha$,3n)Gd$^{145}$.'' The observed half-life of 25(2)~min is in agreement with the currently accepted value of 23.04(4)~min. Previously a 24(1)~min half-life was assigned to either $^{144}$Gd or$^{145}$Gd \cite{1959Olk01}.

\subsection*{$^{146}$Gd}\vspace{0.0cm}
The observation of $^{146}$Gd was reported by Gorddinskii et al. in the 1957 paper ``Neutron-deficient isotopes of the rare earth elements forming as a result of spallation of Ta under bombardment with 660 MeV protons'' \cite{1957Gor01}. A tantalum target was bombarded with 660 MeV protons from the JINR synchrocyclotron. Resultant activities were measured following chromatographic separation. ``We attributed an activity with a period of $\sim$60 days, observed in the Gd fraction to Gd$^{145}$, an isotope not hitherto described in the literature.'' A note added in proof stated ``We are now of the opinion that the Gd and Eu isotopes described in this section actually have mass number 146.'' This half-life is consistent with the currently accepted value of 48.27(10)~d.

\subsection*{$^{147}$Gd}\vspace{0.0cm}
$^{147}$Gd was first observed in 1957 by Shirley et al. as reported in the paper ``Conversion-electron and photon spectra of Gd$^{147}$ and Gd$^{149}$'' \cite{1957Shi01}. An enriched $^{147}$Sm$_{2}$O$_{3}$ target was bombarded with alpha particles from the Berkeley 60-inch cyclotron. Conversion electron spectra were measured with four permanent-magnet 180$^\circ$ spectrographs following chemical separation. ``A new gadolinium isotope decaying by electron capture with a 29-hour half-life was found. Its mass number was determined to be 147 by examination of its excitation function for its production by alpha particle bombardment of Sm$_{2}^{147}$O$_{3}$.'' This half-life is close to the currently accepted value of 38.06(12)h.

\subsection*{$^{148}$Gd}\vspace{0.0cm}
Rasmussen et al. observed $^{148}$Gd in 1953 as described in the paper ``Alpha-radioactivity in the 82-neutron region'' \cite{1953Ras01}. Natural samarium and enriched $^{147}$Sm was bombarded with 38 MeV $\alpha$-particles and europium oxide was bombarded with 50 MeV protons. Resulting $\alpha$-activities were measured with an ionization chamber following chemical separation. ``The mass assignment to 148 rather than to 147 seems more consistent with the long half-life ($>$35 years) of this activity, for the even-odd nucleide Gd$^{147}$ should have considerable energy available for electron capture decay and consequently a much shorter half-life than 35 years.'' The observed half-life limit of $>$35 years is consistent with the currently accepted value of 74.6(3)~y.

\subsection*{$^{149}$Gd}\vspace{0.0cm}
Hoff et al. reported the observation of $^{149}$Gd in the 1951 paper ``Neutron deficient europium and gadolinium isotopes'' in 1951 \cite{1951Hof01}. Natural samarium, enriched $^{147}$Sm$_2$O$_3$ and europium oxide targets were bombarded with $\alpha$-particles and protons with energies between 28 and 36 MeV. Decay curves and absorption spectra were measured following chemical separation. ``The gadolinium isotope, Gd$^{149}$, has been observed as a product in a number of different bombardments. Natural europium oxide was bombarded with protons in a stacked foil arrangement which produced a proton energy range from 32 Mev to 8 Mev. The Gd$^{149}$ was produced in largest yield with 28- to 32-Mev protons, and therefore, a (p,3n) reaction was assumed as the predominant mechanism in the production of this isotope.''. The observed half-life of 9(1)~d is in agreement with the currently accepted value of 9.28(10)~d.

\subsection*{$^{150}$Gd}\vspace{0.0cm}
Rasmussen et al. observed $^{150}$Gd in 1953 as described in the paper ``Alpha-radioactivity in the 82-neutron region'' \cite{1953Ras01}. Eu$_2$O$_3$ was bombarded with 19 MeV deuterons and $^{150}$Gd was formed in the reaction $^{151}$Eu(d,3n). Resulting $\alpha$-activities were measured with an ionization chamber following chemical separation. ``The mass assignment of this alpha-activity to Gd$^{150}$ should be considered tentative, since the assignment is based principally on semi-empirical considerations of the probable expected alpha-decay energy for Gd$^{150}$. Decay measurements over a one-year period show the activity to have a half-life of greater than two years, ruling out the possibility that this activity could arise from l55 day Gd$^{151}$.'' The currently accepted half-life is 1.79(8)~My. We credit Rasmussen et al. with the discovery because their $\alpha$ energy measurement was correct (2.70(15)~MeV) and the assignment was later accepted by Ogawa et al. \cite{1965Oga01}.

\subsection*{$^{151}$Gd}\vspace{0.0cm}
Hein and Voigt published ``Radioactive isotopes of gadolinium'' in 1950 reporting their observation of $^{151}$Gd \cite{1950Hei01}. Eu$_2$O$_3$ targets were bombarded with 20 MeV deuterons from the Berkeley 60-in. cyclotron.  Decay curves, absorption spectra, and $\gamma$-ray spectra were measured following chemical separation. ``The 265-kev gamma-ray must be connected with the decay of Gd$^{151}$. Since this was the only $\gamma$-ray
other than the 102-kev $\gamma$-ray of Gd$^{153}$ the decay scheme of Gd$^{151}$ is probably simple, like that of Gd$^{153}$. The ratio of intensities of the 102-kev $\gamma$-ray to the 265-kev $\gamma$-ray at different times, 590 days apart, was used with the 236-day half-life of the 102-kev $\gamma$-ray to calculate a half-life value of 150 days for Gd$^{151}$.'' This half-life agrees with the currently accepted value of 124(1)~d. Previously a 75~day half-life was assigned to either $^{151}$Gd or $^{153}$Gd with a tentative preference for $^{153}$Gd \cite{1948Kri01}.

\subsection*{$^{152}$Gd}\vspace{0.0cm}
$^{152}$Gd was discovered by Dempster as described in the 1938 paper ``The isotopic constitution of gadolinium, dysprosium, erbium and ytterbium'' \cite{1938Dem01}. Gadolinium oxide reduced with neodymium metal was used for the analysis in the Chicago mass spectrograph. ``A new isotope at mass 154 was observed on four photographs with exposures of ten to seventy minutes, and an isotope at 152 on two plates with seventy minutes exposure.'' Previously Aston had attributed a faint line at 152 to samarium contaminants \cite{1933Ast01}.

\subsection*{$^{153}$Gd}\vspace{0.0cm}
$^{153}$Gd was discovered in 1947 by Inghram et al. as reported in the paper ``Activities induced by pile neutron bombardment of samarium'' \cite{1947Ing02}. A Sm$_{2}$O$_{3}$ sample was irradiated with neutrons at the Hanford Pile. $^{153}$Gd was subsequently identified by mass spectroscopy. ``The 169 activity could conceivably have been caused by samarium, gadolinium, erbium, or ytterbium. Since no line was observed as mass 153 the possibility of samarium was ruled out. Gadolinium, which emits mostly as GdO$^+$, could have caused the line as could erbium or ytterbium emitting as Er$^+$ and Yb$^+$. However, since gadolinium was a known impurity in the samarium sample and no erbium or ytterbium impurities could be detected, it was concluded that the 169 mass line was caused by a gadolinium isotope of mass 153.'' The quoted lower half-life limit of 72~d is consistent with the presently adopted value of 240.4(10)~d.

\subsection*{$^{154}$Gd}\vspace{0.0cm}
$^{154}$Gd was discovered by Dempster as described in the 1938 paper ``The isotopic constitution of gadolinium, dysprosium, erbium and ytterbium'' \cite{1938Dem01}. Gadolinium oxide reduced with neodymium metal was used for the analysis in the Chicago mass spectrograph. ``A new isotope at mass 154 was observed on four photographs with exposures of ten to seventy minutes, and an isotope at 152 on two plates with seventy minutes exposure.'' Previously Aston had attributed a faint line at 154 to samarium contaminants \cite{1933Ast01}.

\subsection*{$^{155-158}$Gd}\vspace{0.0cm}
In the 1933 paper ``Constitution of neodymium, samarium, europium, gadolinium and terbium'' Aston reported the first observation of $^{155}$Gd, $^{156}$Gd, $^{157}$Gd, and $^{158}$Gd \cite{1933Ast01}. Rare earth elements were measured with the Cavendish mass spectrograph. ``Gadolinium (64) appears to consist of 155, 156, 157, 158, and 160. Faint effects at 152, 154, are probably due to the presence of samarium in the sample used.''

\subsection*{$^{159}$Gd}\vspace{0.0cm}
$^{159}$Gd was observed in 1949 by Butement published in the paper ``Radioactive gadolinium and terbium isotopes'' \cite{1949But01}. Natural gadolinium samples were irradiated with neutron in a pile. Gamma- and beta-ray activities were measured following chemical separation. ``The value of $\sigma$ for the 18~hr Gd shows that it cannot be an isomer of Gd$^{161}$, and this activity is presumably, therefore, to be assigned to Gd$^{159}$.'' The measured half-life of 18.0(2)~h agrees with the currently accepted value of 18.479(4)~h. Previously, a 20~h was reported without mass assignments \cite{1947Ser01} and a 18~h  were reported without element or mass assignments and tentatively assigned to $^{161}$Tb \cite{1948Kri01}.

\subsection*{$^{160}$Gd}\vspace{0.0cm}
In the 1933 paper ``Constitution of neodymium, samarium, europium, gadolinium and terbium'' Aston reported the first observation of $^{160}$Gd \cite{1933Ast01}. Rare earth elements were measured with the Cavendish mass spectrograph. ``Gadolinium (64) appears to consist of 155, 156, 157, 158, and 160. Faint effects at 152, 154, are probably due to the presence of samarium in the sample used.''

\subsection*{$^{161}$Gd}\vspace{0.0cm}
$^{161}$Gd was observed in 1949 by Butement published in the paper ``Radioactive gadolinium and terbium isotopes'' \cite{1949But01}. Natural gadolinium samples were irradiated with neutrons in a pile. Gamma- and beta-ray activities were measured following chemical separation. ``The identity, within experimental error, of $\sigma$ for the 218 sec. and the 6.75 d. activities suggests that they must be assigned as follows: $Gd^{161}\, \,{\stackrel{\beta}{\overrightarrow{218 sec.}}}\, \,Tb^{161} \,\,{\stackrel{\beta}{\overrightarrow{6.75 d.}}}\,\,Dy^{161}$stable.'' The observed half-life of 218(5)~s is in agreement with the currently accepted value of 3.646(3)~min.

\subsection*{$^{162}$Gd}\vspace{0.0cm}
Wahlgren et al. reported the observation of $^{162}$Gd in the 1967 paper ``Decay of 10.4-min Gd$^{162}$'' \cite{1967Wah01}. Enriched $^{160}$Gd samples were irradiated with neutrons at the Savannah River high-flux reactor and $^{162}$Gd was produced by double neutron capture. Following chemical separation, $\gamma$- and $\beta$-rays were measured with a NaI(Tl) detector and a Pilot B scintillator assembly, respectively. ``[The figure] shows a comparison of the separated gadolinium and terbium fractions. Concurrent beta-decay curves permit an estimate of the upper limit of branching to 2-h Tb$^{162}$ of $<$2\%. Least-squares analysis of the beta-decay curves gave a half-life for Gd$^{162}$ of 10.4$\pm$0.2~min.'' This half-life agrees with the currently adopted value of 8.4(2)~min.

\subsection*{$^{163}$Gd}\vspace{0.0cm}
``A new isotope $^{163}$Gd; comments on the decay of $^{162}$Gd'' was published in 1982 by Gehrke et al. documenting the observation of $^{163}$Gd \cite{1982Geh02}. $^{163}$Gd was produced in spontaneous fission of $^{252}$Cf and $\gamma$-ray spectra were measured with a Ge(Li) spectrometer following chemical separation. ``A new isotope, $^{163}$Gd, has been identified which was produced in the spontaneous fission of $^{252}$Cf. The half-life of this isotope was measured to be (68$\pm$3)~s and eleven $\gamma$ rays have been assigned to its decay. The assignment of this activity to $^{163}$Gd is based on the presence of these $\gamma$ rays in the gadolinium fraction, which was chemically separated from mixed fission products, and the observation of the growth and decay curve associated with $\gamma$ rays from 19.5-min $^{163}$Tb, the daughter activity.'' This half-life corresponds to the presently accepted value.

\subsection*{$^{164}$Gd}\vspace{0.0cm}
Greenwood et al. reported the discovery of $^{164}$Gd in the 1988 paper ``Identification of a new isotope, $^{164}$Gd'' \cite{1988Gre01}. $^{164}$Gd was produced in spontaneous fission of $^{252}$Cf and $^{164}$Gd was separated with the INEL ISOL facility. X-rays and $\gamma$-rays were measured with an intrinsic planar Ge and Ge(Li) detector, respectively. ``The half-life values in [the table] derived from each of the four measurements, using both mass-l64 and mass-180 fractions, form a consistent set and clearly indicate that in each case the decay of the $^{164}$Gd activity was observed. Thus, the adopted value given in [the table] of (45$\pm$3)~s for the half-life of $^{164}$Gd was obtained straightforwardly as a 1/$\sigma^2$ weighted average of the four individually measured half-life values.'' This half-life corresponds to the presently adopted value.

\subsection*{$^{165}$Gd}\vspace{0.0cm}
$^{165}$Gd was observed in 1998 by Ichikawa et al. as described in ``Identification of $^{161}$Sm and $^{165}$Gd'' \cite{1998Ich01}. $^{238}$U targets were bombarded with 20 MeV protons from the JAERI tandem accelerator facility. $^{165}$Gd was separated with a gas-jet coupled thermal ion source system in the JAERI-ISOL. Beta- and gamma-rays were measured with plastic scintillator and Ge detectors, respectively. ``[The figure] shows the decay curves of the $\beta$-coincident Tb K$_{\alpha 1 + \alpha 2}$ x rays and Tb K$_{\beta 1 + \beta 2}$ + 50.3 keV $\gamma$ ray counts. The respective half-lives were measured to be 11.2$\pm$2.3~s and 9.3$\pm$2.3~s. The average of these two gives the half-life of the $\beta$ decay of $^{165}$Gd, 10.3$\pm$1.6~s.'' This half-life corresponds to the currently accepted value.

\subsection*{$^{166}$Gd}\vspace{0.0cm}
Ichikawa et al. reported the first observation of $^{166}$Gd in ``$\beta$-decay half-lives of new neutron-rich rare-earth isotopes $^{159}$Pm, $^{162}$Sm, and $^{166}$Gd'' in 2005 \cite{2005Ich01}. $^{238}$U targets were bombarded with 15.5 MeV protons from the JAERI tandem accelerator facility. $^{166}$Gd was separated with a gas-jet coupled thermal ion source system in the JAERI-ISOL. Beta- and X/gamma-rays were measured with a sandwich-type plastic scintillator and two Ge detectors, respectively. ``Thus, it was found that these $\gamma$ rays were attributed to the $\beta^-$ decay of $^{166}$Gd. The half-life of $^{166}$Gd was determined to be 4.8$\pm$1.0~s as a weighted average.'' This half-life corresponds to the presently adopted value.

\section{Discovery of $^{135-168}$Tb}

Thirty-one terbium isotopes from A = 135--168 have been discovered so far; these include 1 stable 21 neutron-deficient and 9 neutron-rich isotopes. $^{135}$Tb is the most proton-rich terbium isotope reported, however, the neighboring isotopes $^{136-138}$Tb which are closer to stability have not yet been observed; they could be either bound or unbound with respect to proton emission. According to the HFB-14 model \cite{2007Gor01}, $^{216}$Tb should be the last odd-odd particle stable neutron-rich nucleus while the odd-even particle stable neutron-rich nuclei should continue through $^{219}$Tb. The proton dripline has most likely been reached with the observation that $^{135}$Tb is a proton emitter. Three more isotopes ($^{132-134}$Tb) could possibly still have half-lives longer than 10$^{-9}$~ns \cite{2004Tho01}. Thus, about 56 isotopes have yet to be discovered corresponding to 65\% of all possible promethium isotopes.

\begin{figure}
	\centering
	\includegraphics[scale=.7]{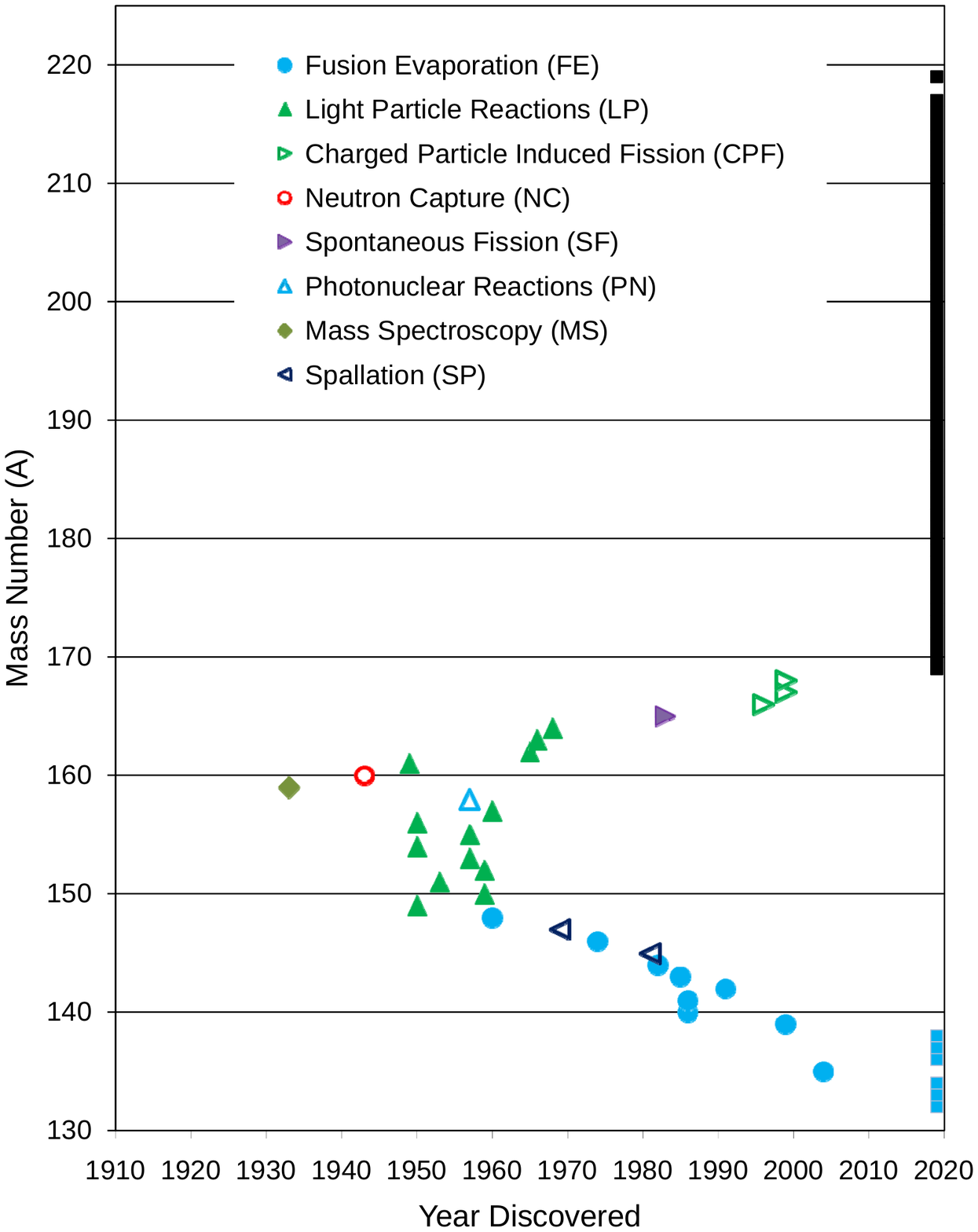}
	\caption{Terbium isotopes as a function of time when they were discovered. The different production methods are indicated. The solid black squares on the right hand side of the plot are isotopes predicted to be bound by the HFB-14 model. On the proton-rich side the light blue squares correspond to unbound isotopes predicted to have lifetimes larger than $\sim 10^{-9}$~s.}
\label{f:year-tb}
\end{figure}

Figure \ref{f:year-tb} summarizes the year of discovery for all terbium isotopes identified by the method of discovery. The range of isotopes predicted to exist is indicated on the right side of the figure. The radioactive terbium isotopes were produced using fusion evaporation reactions (FE), light-particle reactions (LP), charged-particle induced fission (CPF), spontaneous fission (SF), neutron capture (NC), photo-nuclear reactions (PN), and spallation (SP). The stable isotope were identified using mass spectroscopy (MS). Light particles also include neutrons produced by accelerators. In the following, the discovery of each terbium isotope is discussed in detail.

\subsection*{$^{135}$Tb}\vspace{0.0cm}
``Proton decay of the highly deformed nucleus $^{135}$Tb'' was published in 2004 by Woods et al. documenting the discovery of $^{135}$Tb \cite{2004Woo01}. A $^{92}$Mo target was bombarded with a 310 MeV $^{50}$Cr beam from the Argonne ATLAS accelerator system and $^{135}$Tb was formed in the (1p6n) fusion-evaporation reaction. Recoil products were separated with the Fragment Mass Analyzer FMA and protons were measured with a double-sided silicon strip detector. ``[The figure], corresponding to decays occurring within 6 ms of implantation, shows clear evidence for a peak around 1.2 MeV, with very little background present. The energy is too low to be from an $\alpha$-decay so we assign this to proton decay. The energy of the peak is found to be E$_p$=1179(7)~keV using the known ground-state proton decay of $^{147}$Tm as a calibration. This corresponds to a proton decay Q-value Q$_p$=1188(7)~keV. The half-life of the peak is found to be t$_{1/2}$=0.94$_{-0.22}^{+0.33}$~ms using the method of maximum likelihood, and its production cross-section is $\sim$3~nb.'' This corresponds to the currently accepted half-life.

\subsection*{$^{139}$Tb}\vspace{0.0cm}
Xie et al. reported the first observation of $^{139}$Tb in the 1999 paper ``New nuclide $^{139}$Tb and (EC+$\beta^+$) decay of $^{138,139}$Gd'' \cite{1999Xie01}. An enriched $^{106}$Cd target was bombarded with a 220 MeV $^{36}$Ar beam from the Lanzhou cyclotron and $^{139}$Tb was formed in the fusion-evaporation reaction $^{106}$Cd($^{36}$Ar,1p2n). X-rays and $\beta$-delayed $\gamma$-rays were measured in combination with a He-jet tape transport system. ``In the $\gamma$ spectrum gated by Gd-K$_\alpha$ X rays two new $\gamma$ rays with the energies of 109.0- and 119.7-keV were observed. Comparing the excitation functions of the two $\gamma$ rays with that of the 328.4-keV $\gamma$ ray, the most intense $\gamma$ ray of $^{140}$Tb, we assigned the 109.0- and 119.7-keV $\gamma$ rays to the decay of $^{139}$Tb.'' The quoted half-life of 1.6(2)~s corresponds to the presently adopted value.

\subsection*{$^{140,141}$Tb}\vspace{0.0cm}
The first observation of $^{140}$Tb and $^{141}$Tb was reported by Wilmarth et al. in their 1986 paper entitled ``Beta-delayed proton emission in the lanthanide region'' \cite{1986Wil01}. $^{54}$Fe beams of 298 and 276~MeV from the Berkeley SuperHILAC bombarded a $^{92}$Ru target. $^{140}$Tb was produced in the (3p3n) fusion-evaporation reaction and $^{141}$Tb was populated by $\beta$-decay from $^{141}$Dy produced in a (2p3n) reaction. Beta-delayed particles, X-rays and $\gamma$-rays were measured following mass separation with the on-line isotope separator OASIS. ``Delayed protons with a half-life of 2.4$\pm$0.5~s are associated with the new isotope $^{140}$Tb based on coincident Gd K x-rays. The deacy of the 329 keV (2$^{+}$ to 0$^{+}$) $\gamma$-ray following $\beta$-decay of $^{140}$Tb yielded a half-life of 2.3$\pm$0.6~s... From an analysis of the $\gamma$-rays populated in $\beta$-decay, a half-life of 20$\pm$4~s for $^{141}$Gd is in agreement with the $\beta$-delayed proton decay data and the value in [the reference]. This $\gamma$-ray analysis also resulted in the identification of the new isotope $^{141}$Tb with a half-life of 3.5$\pm$0.9~s.'' These half-lives agree with the presently accepted values of 2.4(2)~s and 3.5(2)~s for $^{140}$Tb and $^{141}$Tb, respectively.

\subsection*{$^{142}$Tb}\vspace{0.0cm}
``Decay studies of neutron deficient nuclei near the Z=64 subshell: $^{142}$Dy, $^{140,142}$Tb, $^{140,142}$Gd, $^{140,142}$Eu, $^{142}$Sm and $^{142}$Pm.'' was published in 1991 by Firestone et al. describing the observation of $^{142}$Tb. \cite{1991Fir01}. A $^{92}$Mo metal foil was bombarded with 261 MeV $^{54}$Fe and 224 Mev $^{52}$Cr beams from the Berkeley SuperHILAC accelerator. Reaction products were separated with the OASIS on-line facility and charged particles, $\gamma$-rays, X-rays and positrons were measured. ``We have assigned 18 $\gamma$ rays to the decay of $^{142}$Tb$^g$. The $\gamma$-ray energies and intensities are summarized in [the table]... The intense 515.3-keV transition (resolved from the 511-keV annihilation peak using SAMPO) was analyzed as a two-component parent-daughter decay to extract a 597(17)~ms half-life for $^{142}$Tb$^g$ decay.'' Also, the half-life of an isomeric state was measured to be 303(17)~ms. Both half-lives corresponds to the currently accepted values.

\subsection*{$^{143}$Tb}\vspace{0.0cm}
In 1985 Ollivier et al. reported the discovery of $^{145}$Tb in ``Identification and decay of 12s $^{143}$Tb'' \cite{1985Oll01}. A 191 MeV $^{35}$Cl beam from the Grenoble SARA accelerator bombarded an enriched $^{112}$Sn target and $^{143}$Tb was produced in the fusion-evaporation reaction $^{112}$Sn($^{35}$Cl,2p2n). Gamma-ray coincidences were measured with intrinsic Ge detectors in combination with a He-jet system. ``The $\gamma$-rays determined from $\gamma-\gamma$ coincidence relations are ascribed to the decay of $^{143}$Tb on the basis of following experimental evidences: i) all these $\gamma$-rays are in coincidence with K$_{\alpha_1}$ and K$_{\alpha_2}$ X-rays of Gd. ii) the average measured half-life of the most intensive $\gamma$-rays is found to be 12$\pm$1~s.'' This half-life is the currently accepted value.

\subsection*{$^{144}$Tb}\vspace{0.0cm}
$^{144}$Tb was first observed in 1982 by Sousa et al. published in ``Identification of $^{145}$Tb and $^{144}$Tb and levels in the N=81 nucleus, $^{145}$Gd'' \cite{1982Sou01}. Enriched $^{144}$Sm oxide targets were bombarded with a 129 MeV $^{10}$B from the Texas A\&M isochronous cyclotron and $^{144}$Tb was formed in the reaction $^{144}$Sm($^{10}$B,$\alpha$6n). Gamma-ray spectra were measured with Ge(Li) detectors in combination with a He-jet system. ``One of the latter peaks, 742.9 keV, is assigned to the new isotope $^{144}$Tb. The assignment is based mainly on two facts: (1)  the $\gamma$ ray is in coincidence with gadolinium K x rays, and (2) its energy corresponds to the known 2$^{+}$ (first excited state) $\rightarrow$ 0$^{+}$ (ground state) transition in $^{144}$Gd. In addition, the $\gamma$ ray was first observed at a $^{10}$B energy of 114 MeV, that is, approximately 20-25 MeV above the threshold for the production of $^{145}$Tb.'' The observed half-life of 5(1)~s is in agreement with the currently accepted half-life of 4.25(15)~s. Five month later a 4.5(5)~s half-life for $^{144}$Tb was independently reported \cite{1982Nol02}.

\subsection*{$^{145}$Tb}\vspace{0.0cm}
Alkhazov et al. reported the discovery of $^{145}$Tb in their 1981 paper ``New isotope $^{145}$Tb'' \cite{1981Alk02}. A 1 GeV proton beam bombarded a tungsten target at the Leningrad Nuclear Physics Institute. $^{145}$Tb was formed in spallation reactions and identified with the IRIS on-line mass-separator facility. X-ray and $\gamma$-ray spectra were measured with Ge(Li) detectors. ``On the basis of these facts the transitions with half-life T$_{1/2}$ = 29.5$\pm$1.5~s have been assigned to the decay of $^{145}$Tb.'' This half-life agree with the presently adopted value of 30.9(7)~s.

\subsection*{$^{146}$Tb}\vspace{0.0cm}
$^{146}$Tb was first observed in 1974 by Newman et al. as reported in `` Levels in $^{146,147,148}$Gd observed following the decay of terbium parents including a new isotopes, $^{146}$Tb'' \cite{1974New02}. A 118~MeV $^{12}$C beam from the Oak Ridge isochronous cyclotron bombarded a $^{141}$Pr target. Gamma-ray singles and coincidences were measured with Ge(Li) detectors. ``The assignment of the new 23-sec activity to $^{146}$Tb is based primarily on the fact that five of its $\gamma$-rays have been observed by Kownacki et al. in a $^{144}$Sm($\alpha$,2n$\gamma$) study.'' This half-life agrees with the currently accepted value of 24.1(5)~s.

\subsection*{$^{147}$Tb}\vspace{0.0cm}
In the 1969 paper ``Isomerism in $^{147}$Tb'' Chu et al. described the identification of $^{147}$Tb \cite{1969Chu01}. Tantalum targets were irradiated with a 28~GeV protons from the Brookhaven Alternating Gradient Synchrotron. Subsequent activities were measured with NaI(Tl) and Ge(Li) detectors following chemical separation. ``Two $^{147}$Tb isomers with half-lives of 1.1$\pm$0.17 h and 2.5$\pm$0.1 min were found and their $\gamma$-ray spectra determined.'' These half-lives are in agreement with the currently accepted values of 1.64(3)~h for the ground state and 1.87(5)~min the isomeric state. A previously measured 24 min half-life tentatively assigned to $^{147}$Tb \cite{1960Tot01} was evidently incorrect.

\subsection*{$^{148}$Tb}\vspace{0.0cm}
Toth et al. discovered $^{148}$Tb as reported in ``Two new neutron-deficient Terbium isotopes'' in 1960 \cite{1960Tot01}. A $^{141}$Pr target was bombarded with a 110 MeV $^{14}$N and 65 and 75~MeV $^{12}$C beams from the Berkeley heavy-ion linear accelerator. Gamma-ray spectra were measured following chemical separation. ``The two $\gamma$-rays have thus been assigned to $^{148}$Tb decay, not only on the basis of the variation of their intensities with bombarding energies, but also because of their half-lives.'' The observed half-life of 70~min is consistent with the currently adopted value of 60(1)~min.

\subsection*{$^{149}$Tb}\vspace{0.0cm}
$^{149}$Tb was observed in 1950 by Rasmussen et al. and published in the paper ``Mass assignments of alpha-active isotopes in the rare-earth region'' \cite{1950Ras01}. Gadolinium oxide targets were bombarded with 150 MeV protons from the Berkeley 184-inch cyclotron. $^{149}$Tb was identified with a mass spectrograph following chemical separation. ``The mass assignment of the alpha-emitting terbium isotope of 4.0-hr. half-life and 4.0-Mev alpha-particle energy was made by performing a mass spectrographic separation of terbium activity onto a photographic plate and detecting alpha-activity by a transfer plate technique... A concentration of alpha-tracks was observed on the transfer plate only in a region corresponding to mass 149. The 4-hr. terbium alpha-activity was the only alpha-activity here present in large enough amount to be detected by this technique.'' This half-life agrees with the presently accepted values of 4.118(25)~h.

\subsection*{$^{150}$Tb}\vspace{0.0cm}
``$^{150}$Tb: A new terbium isotope'' was published in 1959 by Toth et al. documenting their observation of $^{150}$Tb \cite{1959Tot01}.  The Uppsala synchrocyclotron was used to bombard natural gadolinium targets with 60 MeV protons to produce $^{150}$Tb. Decay curves and $\gamma$-ray spectra were measured following mass and chemical separation. ``Four decay curves were obtained counting the mass 150 samples `A' and `B' in a flow-type proportional counter and in a single channel scintillation spectrometer. The decay curves when resolved, yielded the same half-lives, i.e., 3.1$\pm$0.2 hr (Tb$^{150}$), 17.5$\pm$0.5 hr (Tb$^{151}$) as well as a longer half-life that was found to be a general background present in all the masses.'' This half-life agrees with the currently accepted value of 3.48(16)~h.

\subsection*{$^{151}$Tb}\vspace{0.0cm}
Rasmussen et al. observed $^{151}$Tb in 1953 as described in the paper ``Alpha-radioactivity in the 82-neutron region'' \cite{1953Ras01}. Eu$_2$O$_3$ and gadolinium targets were bombarded with 45 MeV $\alpha$-particles and 100 MeV protons to form $^{150}$Gd in the reactions $^{151}$Eu($\alpha$,4n) and Gd(p,xn), respectively. Resulting $\alpha$-activities were measured with an ionization chamber following chemical separation. ``Only a tentative mass assignment of the 19-hour activity can be made at present. The alpha-particle excitation work on europium oxide by Roller and Rasmussen indicated a probable mass assignment to 151, with 150 a possibility.'' The observed half-life of 19(1)~h agrees with the presently adopted value of 17.609(1)~h. The reference of Roller and Rasmussen mentioned in the quote were unpublished results.

\subsection*{$^{152}$Tb}\vspace{0.0cm}
Toth et al. discovered $^{152}$Tb in the 1959 article ``New terbium isotope, Tb$^{152}$'' \cite{1959Tot02}. Enriched $^{151}$Eu and $^{153}$Eu targets were bombarded with 37 and 48~MeV $\alpha$-particles from the Berkeley 60-in. cyclotron. Gamma-ray spectra were measured with NaI(Tl) detectors following chemical separation. ``A careful examination of photon and electron spectra from cyclotron-produced mixtures of light terbium isotopes has led to the identification of a new isotope, Tb$^{152}$, with an 18.5-hr half-life.'' This half-life is in agreement with the currently accepted value of 17.5(1)~h. Four months later a 4.0(5)~min half-life corresponding to an isomeric state of $^{152}$Tb was measured independently \cite{1959Olk02}. A 4.5~h half-life previously assigned to $^{152}$Tb \cite{1948Wil02} was evidently incorrect.

\subsection*{$^{153}$Tb}\vspace{0.0cm}
``Nuclear spectroscopy of neutron-deficient rare earths (Tb through Hf)'' was published in 1957 by Mihelich et al. describing the observation of $^{153}$Tb \cite{1957Mih01}. Different enriched rare earth elements were irradiated with 12$-$22 MeV protons from the ORNL 86-inch cyclotron. The resulting activities were measured with a conversion electron spectrograph and a scintillation counter following chemical separation. ``Tb$^{153}$(62~hr)$\rightarrow$Gd$^{153}$: The presence of this activity is established by observation of the daughter activity, Gd$^{153}$, which decays to levels in Eu$^{153}$. This conclusion is consistent with the yields from various mass-enriched targets'' This half-life is in agreement with the currently accepted value of 2.34(1)~d. A previously reported half-life of 5.1~d \cite{1948Wil02,1950Wil03} was evidently incorrect.

\subsection*{$^{154}$Tb}\vspace{0.0cm}
Wilkinson and Hicks published the observation of $^{154}$Tb in the 1950 paper ``Radioactive isotopes of the rare earths. III. Terbium and holmium isotopes'' \cite{1950Wil03}. Europium and gadolinium targets were bombarded with $\alpha$-particles and protons, respectively, from the Berkeley 60-in. cyclotron. Electrons, positrons, $\gamma$-rays and X-rays were measured following chemical separation. ``17.2$\pm$0.2-Hr. Tb$^{154}$: In all bombardments of europium with $\alpha$-particles, and in low yields in proton bombardments of gadolinium a 17.2- hr. positron emitting activity was observed.'' This half-life is close to the currently adopted values of 21.54(4)~h for the ground state and 22.7(5)~h for an isomeric state. Wilkinson and Hicks had assigned this half-life to $^{154}$Tb two years earlier, however, at that time they classified it as ``D'' which means that they were only sure about the element and not the mass assignment. We credit Wilkinson and Hicks with the discovery although they incorrectly identified the neighboring nuclei $^{153}$Tb and $^{155}$Tb because subsequently their results were not questioned \cite{1955Han02,1957Mih01}.

\subsection*{$^{155}$Tb}\vspace{0.0cm}
``Nuclear spectroscopy of neutron-deficient rare earths (Tb through Hf)'' was published in 1957 by Mihelich et al. describing the observation of $^{155}$Tb \cite{1957Mih01}. Different enriched rare earth elements were irradiated with 12$-$22 MeV protons from the ORNL 86-inch cyclotron. The resulting activities were measured with a conversion electron spectrograph and a scintillation counter following chemical separation.  ``Tb$^{155}$(5.6~days)$\rightarrow$Gd$^{155}$: ... By superimposing spectrograms, one is able to make a very sensitive test as to the `identicality' of sets of conversion lines. Hence, it was evident that we were indeed producing Tb$^{155}$. To confirm our results, we irradiated two targets with 12-MeV protons to produce the ($4p,n$) reaction alone; the targets were enriched in masses 155 and 156, respectively. Although neither target was enriched to a high degree, the relative intensity of the Tb$^{155}$ and Tb$^{156}$ transitions were consistent with the isotopic enrichment factors.'' This half-life agrees with the presently adopted value of 5.32(6)~d. Previously reported half-lives of $\sim$1~y \cite{1948Wil02} and 190(5)~days \cite{1950Wil03} were evidently incorrect.

\subsection*{$^{156}$Tb}\vspace{0.0cm}
Wilkinson and Hicks published the observation of $^{156}$Tb in the 1950 paper ``Radioactive isotopes of the rare earths. III. Terbium and holmium isotopes'' \cite{1950Wil03}. Europium targets were bombarded with 19 MeV $\alpha$-particles from the Berkeley 60-in. cyclotron. Electrons, positrons, $\gamma$-rays and X-rays were measured following chemical separation. ``5.0$\pm$0.1-Hr. Tb$^{156}$: In bombardments of europium with 19-MeV $\alpha$-particles, an activity of half-life 5.0$\pm$0.1 hr. measured through nine half-lives was observed in high yield.'' This half-life agrees with the currently accepted half-life of 5.3(2)~h for an isomeric state. We credit Wilkinson and Hicks with the discovery although they incorrectly identified the neighboring nuclei $^{155}$Tb and $^{157}$Tb because subsequently their results were not questioned \cite{1955Han02,1957Mih01}.

\subsection*{$^{157}$Tb}\vspace{0.0cm}
$^{157}$Tb was first observed in 1960 by Naumann et al. as reported in ``Preparation of long-live terbium-157 and terbium 158'' \cite{1960Nau01}. Dysprosium oxide enriched in $^{156}$Dy was bombarded with neutrons in the materials testing reactor. $^{157}$Tb was identified with a mass spectrometer following chemical separation. ``Mass-spectrometric investigation of the terbium fraction revealed the existence of mass peaks at 157, 158, 159 and 160, indicating the preparation of new long-lived isotopes terbium-157 and -158 in addition to the stable terbium-159 and the 76-day terbium-160.''  The quoted half-life limit of $>$30~y is consistent with the currently accepted value of 71(7)~y. Previously only limits of $<$30~min or $>$100~y \cite{1953Han02}, $<$10~min or $>$5~y \cite{1955Han02} and $<$4~h or $>$10~y \cite{1960Tot02}. Also, an earlier report of a 4.7(1)~d half-life \cite{1950Wil03} was evidently incorrect.

\subsection*{$^{158}$Tb}\vspace{0.0cm}
Hammer and Stewart published the observation of $^{158}$Tb in the 1957 paper ``Isomeric transitions in the rare-earth elements'' \cite{1957Ham01}. Terbium oxides were irradiated with x-rays from the 75 MeV Iowa State College synchrotron and $^{158}$Tb was produced in the photonuclear ($\gamma$,n) reaction. Decay curves and X- and $\gamma$-ray spectra were recorded. ``Since Tb is a single isotope of mass 159, the isomeric transition therefore occurs in Tb$^{158}$.''  The observed half-life of 11.0(1)~s agrees with the currently accepted value of 10.70(17)~s for an isomeric state. A previously assigned 3.6~min half-life \cite{1938Poo02} was evidently incorrect.

\subsection*{$^{159}$Tb}\vspace{0.0cm}
In the 1933 paper ``Constitution of neodymium, samarium, europium, gadolinium and terbium'' Aston reported the first observation of $^{159}$Tb \cite{1933Ast01}. Rare earth elements were measured with the Cavendish mass spectrograph. ``Terbium shows only one line, 159. Search was made for a possible heavier constituent suggested by its atomic weight, 159.2, but none could be detected.''

\subsection*{$^{160}$Tb}\vspace{0.0cm}
The discovery of $^{160}$Tb was reported by Bothe in the 1943 paper ``Eine langlebige Terbium-Aktivit\"at'' \cite{1943Bot01}. A terbium oxide sample was irradiated with thermal neutrons produced by deuteron bombardment of beryllium. Decay curves and absorption spectra were measured. ``Nach den bisherigen Messungen, die sich \"uber 6 Monate erstrecken, betr\"agt die Halbwertszeit 72$\pm$3~d... Da nur die Reaktion Tb$^{159}$(n,$\gamma$)Tb$^{160}$ in Frage kommt, geh\"ort die neue Aktivit\"at dem Tb$^{160}$ an.'' [According the measurements up to now, which lasted for 6 months, the half-life is 72$\pm$3~d... Because only the reaction Tb$^{159}$(n,$\gamma$)Tb$^{160}$ is possible, this activity belongs to Tb$^{160}$.] This half-life agrees with the presently adopted value of 72.3(2)~d. Previously assigned half-lives of 3.9~h \cite{1936Hev01} and 3.3~h \cite{1938Poo02} were evidently incorrect. Sugden had reported a 3.9(1)~h half-life without a mass assignment \cite{1935Sug01}.

\subsection*{$^{161}$Tb}\vspace{0.0cm}
$^{161}$Tb was observed in 1949 by Butement published in the paper ``Radioactive gadolinium and terbium isotopes'' \cite{1949But01}. Natural gadolinium samples were irradiated with neutrons in a pile. Gamma- and beta-ray activities were measured following chemical separation. ``The identity, within experimental error, of $\sigma$ for the 218 sec. and the 6.75 d. activities suggests that they must be assigned as follows: $Gd^{161}\, \,{\stackrel{\beta}{\overrightarrow{218 sec.}}}\, \,Tb^{161} \,\,{\stackrel{\beta}{\overrightarrow{6.75 d.}}}\,\,Dy^{161}$ stable.'' The observed half-life of 6.75~d is in agreement with the currently accepted value of 6.906(19)~d. Previously, a 18~h was reported without element or mass assignments and tentatively assigned to $^{161}$Tb \cite{1948Kri01}.

\subsection*{$^{162}$Tb}\vspace{0.0cm}
Schneider and M\"unzel reported the observation of $^{162}$Tb in the 1965 paper ``Notiz zum Zerfall von $^{162}$Tb'' \cite{1965Sch01}. The Karlruhe synchrocyclotron was used to bombard an enriched $^{162}$Dy$_2$O$_3$ target with 30 MeV deuterons. Beta- and gamma-rays were measured with a methan flow counter and a NaI(Tl) crystal, respectively. ``Bei Vorgabe von 15.8 min f\"ur die $^{163}$Tb-Verunreinigung lieferte die Computer-Analyse der $\gamma$-Abfallskurven 7.43$\pm$0.04 min (keine 2 h-Aktivit\"at); dagegen lieferte die Analyse der $\beta$-Abfallskurve zwei Komponenten mit Halbwertszeiten von 7.48$\pm$0.03 und 134$\pm$1 min.'' [With a fixed half-life for the $^{163}$Tb impurities the computer analysis of the $\gamma$-decay curves gave  7.43$\pm$0.04 min (no 2~h activity); in contrast the analysis of the $\beta$-decay curve resulted in two components with half-lives of 7.48$\pm$0.03 and 134$\pm$1 min.] The 7.4~min half-life agrees with the presently adopted value of 7.60(15)~min. The other state (134~min) was not correct. A $\sim$8~min was mentioned earlier \cite{1962Tak01} referring to a paper ``to be published''.

\subsection*{$^{163}$Tb}\vspace{0.0cm}
In the 1966 paper ``Der Zerfall von $^{162}$Tb und $^{163}$Tb'' Funke et al. identified $^{163}$Tb \cite{1966Fun01}. Dysprosium oxide enriched in $^{164}$Dy was irradiated with the 27 MeV Jena betatron and $^{163}$Tb was produced in the photonuclear ($\gamma$,p) reaction. Gamma- and beta-spectra were measured with Ge(Li) and scintillation detectors. ``Aus dem Intensit\"atsabfall des integralen $\beta$- und $\gamma$-Spektrums sowie dem Abfall einzelner Linien wurde die Halbwertszeit von $^{163}$Tb zu T$_{1/2}$ = 19.5$\pm$0.5~min bestimmt.'' [From the intensity decrease of the integral $\beta$- and $\gamma$- spectra as well as the decay of individual lines the half-life was determined to be 19.5$\pm$0.5~min.] This half-life agrees with the currently adopted value of 19.5(3)~min. Previously reported half-lives of 7(1)~min \cite{1960Wil02}, 6.5(3)~y \cite{1960Als01}, and 6.5~y \cite{1962Tak01}, were evidently incorrect.

\subsection*{$^{164}$Tb}\vspace{0.0cm}
$^{164}$Tb was discovered by Monnand and Moussa in the 1968 paper ``D\'esint\'egration du terbium 164'' \cite{1968Mon01}. An enriched $^{164}$Dy target was irradiated with 14 MeV neutrons at the Grenoble 400 kV Sames accelerator and $^{164}$Tb was produced in (n,p) charge exchange reactions. Decay curves and $\gamma$-ray spectra were measured with a Ge(Li) detector. ``Nous avons mesur\'e la p\'eriode de $^{164}$Tb en suivant la d\'ecroissance des principales raies (169, 215, 611, 618 et 754 keV), nous avons obtenu une p\'eriode: T$_{1/2}$ = 3.2$\pm$0.2 mn, en bon accord avec celle indiqu\'ee par Kaffrell et Herrmann: 3.04$\pm$0.03 mn.'' [We measured the half-life of $^{164}$Tb from the decay of the dominant lines (169, 215, 611, 618, and 754 keV) to be T$_{1/2}$ = 3.2$\pm$0.2 min, in good agreement with the one indicated by Kaffrell and Herrmann: 3.04$\pm$0.03 min.] This half-life agrees with the presently adopted value of 3.0(1)~min. The reference of Kaffrell and Herrmann was published in an conference abstract.

\subsection*{$^{165}$Tb}\vspace{0.0cm}
In the article ``Identification of a new isotope, $^{165}$Tb'' Greenwood et al. discovered $^{165}$Tb in 1983 \cite{1983Gre01}. $^{165}$Tb was produced in spontaneous fission of $^{252}$Cf and $\gamma$-ray spectra were measured with a Ge(Li) spectrometer following chemical separation. ``A previously unreported isotope, $^{165}$Tb, has been identified, and its half-life has been measured
to be 2.11$\pm$0.10 min.'' This half-life corresponds to the currently accepted value.

\subsection*{$^{166}$Tb}\vspace{0.0cm}
Ichikawa et al. published the discovery of $^{166}$Tb in the 1996 paper ``Mass separation of neutron-rich isotopes using a gas-jet coupled thermal ion source'' documenting their observation of $^{166}$Tb \cite{1996Ich01}. A $^{238}$U target was bombarded with 15 MeV protons from the JAERI tandem accelerator. Fission fragments were separated by mass with a gas-jet coupled to a thermal ion source and identified by measuring $\gamma$-ray spectra. ``The first observation of a new isotope $^{166}$Tb was carried out using its monoxide ions from this ion source: T$_{1/2}$($^{166}$Tb) = 21$\pm$6~s.'' This value agrees with the currently accepted value of 25.6(22)~s.

\subsection*{$^{167,168}$Tb}\vspace{0.0cm}
``$\beta$-decay half-lives of new neutron-rich isotopes $^{167,168}$Tb and levels in $^{167,168}$Dy'' was published in 1999 by Asai et al. documenting the observation of $^{167}$Tb and $^{168}$Tb. A $^{238}$U target was bombarded with 20 MeV protons from the JAERI tandem accelerator. Fission fragments were separated by mass with a gas-jet coupled to a thermal ion source and identified by measuring $\beta$- and $\gamma$-ray singles and coincidences. ``$\beta$-decay half-lives of new neutron-rich isotopes $^{167}$Tb and $^{168}$Tb have been determined to be 19.4(27)~s and 8.2(13)~s, respectively.'' The $^{167}$Tb half-life agrees with the currently accepted value of 19(3)~s while the $^{168}$Tb half-life corresponds to the present value.

\section{Summary}
The discoveries of the known samarium, europium, gadolinium, and terbium isotopes have been compiled and the methods of their production discussed.
While the identification of the even Z elements samarium and gadolinium was relatively easy, the identification of the odd Z elements europium and terbium proved difficult.

Only two samarium isotopes were at first incorrectly identified ($^{151}$Sm and $^{157}$Sm) and for two others the half-lives were measured without a mass assignment ($^{143}$Sm and $^{155}$Sm). Similarly, in gadolinium, the half-lives of $^{137}$Gd and $^{142}$Gd were initially incorrect and the half-lives of $^{145}$Gd, $^{151}$Gd, and $^{159}$Gd were not immediately assigned.

In contrast, in europium, although only the assignments of $^{152}$Eu and $^{159}$Eu were initially uncertain, nine half-lives were measured incorrectly ($^{137,138}$Eu, $^{140}$Eu, $^{144}$Eu, $^{146,147}$Eu, $^{149,150}$Eu, and $^{160}$Eu). In terbium, also nine isotopes were at first misidentified ($^{147}$Tb $^{152,153}$Tb, $^{155}$Tb, $^{157,158}$Tb, $^{160,161}$Tb and $^{163}$Tb).

\ack

This work was supported by the National Science Foundation under grants No. PHY06-06007 (NSCL).

%%% Here we use thebibliography environment to produce the reference list,
%%% but you can use BibTeX as well:
\bibliography{../isotope-discovery-references}

\newpage

%%% Please start a new page by uncommenting the next
\newpage

\TableExplanation

\bigskip
\renewcommand{\arraystretch}{1.0}

\section{Table 1.\label{tbl1te} Discovery of samarium, europium, gadolinium, and terbium isotopes }
\begin{tabular*}{0.95\textwidth}{@{}@{\extracolsep{\fill}}lp{5.5in}@{}}
\multicolumn{2}{p{0.95\textwidth}}{ }\\

Isotope & Samarium, europium, gadolinium, and terbium  isotope \\
Author & First author of refereed publication \\
Journal & Journal of publication \\
Ref. & Reference \\
Method & Production method used in the discovery: \\

  & FE: fusion evaporation \\
  & LP: light-particle reactions (including neutrons) \\
  & MS: mass spectroscopy \\
  & NC: neutron capture reactions \\
  & PN: photo-nuclear reactions \\
  & NF: neutron induced fission \\
  & CPF: charged-particle induced fission \\
  & SF: spontaneous fission \\
  & SP: spallation \\

Laboratory & Laboratory where the experiment was performed\\
Country & Country of laboratory\\
Year & Year of discovery \\
\end{tabular*}
\label{tableI}

\datatables % This command is necessary to get the table names in toc

%% One-page data tables are also best formatted using the longtable
%% environment:
%\begin{longtable}{c}
%\caption{This is the First Data Table}\\
%\endhead\\
%\end{longtable}

%% If the table is to span over the whole text width, we set:

\setlength{\LTleft}{0pt}
\setlength{\LTright}{0pt}

% To avoid ``Overfull \hboxes...'' decrease the intercolumn spacing:

\setlength{\tabcolsep}{0.5\tabcolsep}

\renewcommand{\arraystretch}{1.0}

\footnotesize % we need to squeeze the font size a lot!

\begin{longtable}{@{\extracolsep\fill}llllllll@{}}
\caption{Discovery of samarium, europium, gadolinium, and terbium isotopes. See page\ \pageref{tbl1te} for Explanation of Tables}
Isotope & Author & Journal & Ref. & Method & Laboratory & Country & Year\\
\hline\\
\endfirsthead\\
\caption[]{(continued)}
Isotope & Author & Journal & Ref. & Method & Laboratory & Country & Year\\
\hline\\
\endhead
$^{129}$Sm & S.-W. Xu & Phys. Rev. C &\cite{1999Xu01}& FE & Lanzhou & China &1999 \\
$^{130}$Sm & A.A. Sonzogni & Phys. Rev. Lett. &\cite{1999Son01}& FE & Argonne & USA &1999 \\
$^{131}$Sm & P.A. Wilmarth & Z. Phys. A &\cite{1986Wil01}& FE & Berkeley & USA &1986 \\
$^{132}$Sm & R. Wadsworth & Z. Phys. A &\cite{1989Wad01}& FE & Daresbury & UK &1989 \\
$^{133}$Sm & D.D. Bogdanov & Nucl. Phys. A &\cite{1977Bog01}& FE & Dubna & Russia &1977 \\
$^{134}$Sm & D.D. Bogdanov & Nucl. Phys. A &\cite{1977Bog01}& FE & Dubna & Russia &1977 \\
$^{135}$Sm & D.D. Bogdanov & Nucl. Phys. A &\cite{1977Bog01}& FE & Dubna & Russia &1977 \\
$^{136}$Sm & M. Nowicki & Acta Phys. Pol. B &\cite{1982Now01}& FE & Dubna & Russia &1982 \\
$^{137}$Sm & N. Redon & Z. Phys. A &\cite{1986Red01}& FE & Grenoble & France &1986 \\
$^{138}$Sm & M. Nowicki & Acta Phys. Pol. B &\cite{1982Now01}& FE & Dubna & Russia &1982 \\
$^{139}$Sm & J. van Klinken & Z. Phys. A &\cite{1971van01}& LP & Karlsruhe & Germany &1971 \\
$^{140}$Sm & E. Herrmann & Radiochim. Acta &\cite{1967Her01}& SP & Dubna & Russia &1967 \\
$^{141}$Sm & E. Herrmann & Radiochim. Acta &\cite{1967Her01}& SP & Dubna & Russia &1967 \\
$^{142}$Sm & I. Gratot & Nucl. Phys. &\cite{1959Gra01}& SP & Orsay & France &1959 \\
$^{143}$Sm & E. Silva & Nuovo Cimento &\cite{1956Sil01} & PN & Sao Paulo & Brazil &1956 \\
$^{144}$Sm & F.W. Aston & Proc. Roy. Soc. A &\cite{1934Ast03}& MS & Cambridge & UK &1933 \\
$^{145}$Sm & M.G. Inghram & Phys. Rev. &\cite{1947Ing02}& NC & Argonne & USA &1947 \\
$^{146}$Sm & D.C. Dunlavey & Phys. Rev. &\cite{1953Dun01}& LP & Berkeley & USA &1953 \\
$^{147}$Sm & F.W. Aston & Nature &\cite{1933Ast01}& MS & Cambridge & UK &1933 \\
$^{148}$Sm & F.W. Aston & Nature &\cite{1933Ast01}& MS & Cambridge & UK &1933 \\
$^{149}$Sm & F.W. Aston & Nature &\cite{1933Ast01}& MS & Cambridge & UK &1933 \\
$^{150}$Sm & F.W. Aston & Proc. Roy. Soc. A &\cite{1934Ast03}& MS & Cambridge & UK &1934 \\
$^{151}$Sm & M.G. Inghram & Phys. Rev. &\cite{1947Ing02}& NC & Argonne & USA &1947 \\
$^{152}$Sm & F.W. Aston & Nature &\cite{1933Ast01}& MS & Cambridge & UK &1933 \\
$^{153}$Sm & M.L. Pool & Phys. Rev. &\cite{1938Poo02}& LP & Michigan & USA &1938 \\
$^{154}$Sm & F.W. Aston & Nature &\cite{1933Ast01}& MS & Cambridge & UK &1933 \\
$^{155}$Sm & L. Winsberg & Nat. Nucl. Ener. Ser. &\cite{1951Win02}& NF & Los Alamos & USA &1951 \\
$^{156}$Sm & L. Winsberg & Nat. Nucl. Ener. Ser. &\cite{1951Win03}& NF & Argonne & USA &1951 \\
$^{157}$Sm & J.M. D'Auria & Can. J. Phys. &\cite{1973DAu01}& LP & Simon Fraser & Canada &1973 \\
$^{158}$Sm & J.B. Wilhelmy & Phys. Rev. Lett. &\cite{1970Wil01}& SF & Berkeley & USA &1970 \\
$^{159}$Sm & H. Mach & Phys. Rev. Lett. &\cite{1986Mac01}& NF & Brookhaven & USA &1986 \\
$^{160}$Sm & H. Mach & Phys. Rev. Lett. &\cite{1986Mac01}& NF & Brookhaven & USA &1986 \\
$^{161}$Sm & S. Ichikawa & Phys. Rev. C &\cite{1998Ich01}& CPF & JAERI & Japan &1998 \\
$^{162}$Sm & S. Ichikawa & Phys. Rev. C &\cite{2005Ich01}& CPF & JAERI & Japan &2005 \\
 & & & & & & & \\
  & & & & & & & \\
$^{130}$Eu & C.N. Davids & Phys. Rev. C &\cite{2004Dav01}& FE & Argonne & USA &2004 \\
$^{131}$Eu & C.N. Davids & Phys. Rev. Lett. &\cite{1998Dav01}& FE & Argonne & USA &1998 \\
$^{132}$Eu  & & & & & & & \\
$^{133}$Eu  & & & & & & & \\
$^{134}$Eu & K.S. Vierinen & Nucl. Phys. A &\cite{1989Vie02}& FE & Berkeley & USA &1989 \\
$^{135}$Eu & K.S. Vierinen & Nucl. Phys. A &\cite{1989Vie02}& FE & Berkeley & USA &1989 \\
$^{136}$Eu & B.D. Kern & Phys. Rev. C &\cite{1987Ker01}& FE & Oak Ridge & USA &1987 \\
$^{137}$Eu & M. Nowicki & Acta Phys. Pol. B &\cite{1982Now01}& FE & Dubna & Russia &1982 \\
$^{138}$Eu & M. Nowicki & Acta Phys. Pol. B &\cite{1982Now01}& FE & Dubna & Russia &1982 \\
$^{139}$Eu & J. van Klinken & Phys. Rev. C &\cite{1975van01}& LP & Groningen & Netherlands &1975 \\
$^{140}$Eu & M. Nowicki & Acta Phys. Pol. B &\cite{1982Now01}& FE & Dubna & Russia &1982 \\
$^{141}$Eu & J. Deslauriers & Z. Phys. A &\cite{1977Des01}& LP & McGill & Canada &1977 \\
$^{142}$Eu & H,P. Malan& Radiochim. Acta &\cite{1966Mal01}& LP & Karlsruhe & Germany &1966 \\
$^{143}$Eu & K. Kotajima & Nucl. Phys. &\cite{1965Kot01}& LP & Amsterdam & Netherlands &1965 \\
$^{144}$Eu & R. Messlinger & Phys. Lett. &\cite{1965Mes01}& LP & Heidelberg & Germany &1965 \\
$^{145}$Eu & R.W. Hoff & Phys. Rev. &\cite{1951Hof01}& LP & Berkeley & USA &1951 \\
$^{146}$Eu & G.M. Gorodinskii & Bull. Acad. Sci. USSR &\cite{1957Gor01}& SP & Dubna & Russia &1957 \\
$^{147}$Eu & R.W. Hoff & Phys. Rev. &\cite{1951Hof01}& LP & Berkeley & USA &1951 \\
$^{148}$Eu & R.W. Hoff & Phys. Rev. &\cite{1951Hof01}& LP & Berkeley & USA &1951 \\
$^{149}$Eu & N.A. Antoneva & Bull. Acad. Sci. USSR &\cite{1959Ant01}& SP & Leningrad & Russia &1959 \\
$^{150}$Eu & F.D.S. Butement & Nature &\cite{1950But01}& PN & Harwell & UK &1950 \\
$^{151}$Eu & F.W. Aston & Nature &\cite{1933Ast01}& MS & Cambridge & UK &1933 \\
$^{152}$Eu & M.L. Pool & Phys. Rev. &\cite{1938Poo02}& LP & Michigan & USA &1938 \\
$^{153}$Eu & F.W. Aston & Nature &\cite{1933Ast01}& MS & Cambridge & UK &1933 \\
$^{154}$Eu & M.G. Inghram & Phys. Rev. &\cite{1947Ing03}& NC & Argonne & USA &1947 \\
$^{155}$Eu & M.G. Inghram & Phys. Rev. &\cite{1947Ing02}& NC & Argonne & USA &1947 \\
$^{156}$Eu & M.G. Inghram & Phys. Rev. &\cite{1947Ing02}& NC & Argonne & USA &1947 \\
$^{157}$Eu & L. Winsberg & Nat. Nucl. Ener. Ser. &\cite{1951Win01}& NF & Argonne & USA &1951 \\
$^{158}$Eu & L. Winsberg & Nat. Nucl. Ener. Ser. &\cite{1951Win01}& NF & Argonne & USA &1951 \\
$^{159}$Eu & T. Kuroyanagi & J. Phys. Soc. Japan &\cite{1961Kur01}& PN & Tohoku & Japan &1961 \\
$^{160}$Eu & J.M. D'Auria & Can. J. Phys. &\cite{1973DAu01}& LP & Simon Fraser & Canada &1973 \\
$^{161}$Eu & H. Mach & Phys. Rev. Lett. &\cite{1986Mac01}& NF & Brookhaven & USA &1986 \\
$^{162}$Eu & R.C. Greenwood & Phys. Rev. C &\cite{1987Gre01}& SF & Idaho Falls & USA &1987 \\
$^{163}$Eu & H. Hayashi & Eur. Phys. J. A &\cite{2007Hay01}& CPF & Tokai& Japan &2007 \\
$^{164}$Eu & H. Hayashi & Eur. Phys. J. A &\cite{2007Hay01}& CPF & Tokai& Japan &2007 \\
$^{165}$Eu & H. Hayashi & Eur. Phys. J. A &\cite{2007Hay01}& CPF & Tokai& Japan &2007 \\
 & & & & & & & \\
 & & & & & & & \\
$^{135}$Gd & S. Xu & Z. Phys. A &\cite{1996Xu01}& FE & Lanzhou & China &1996 \\
$^{136}$Gd  & & & & & & & \\
$^{137}$Gd & S.-W. Xu & Phys. Rev. C &\cite{1999Xu01}& FE & Lanzhou & China &1999 \\
$^{138}$Gd & C.J. Lister & Phys. Rev. Lett. &\cite{1985Lis01}& FE & Daresbury & UK &1985 \\
$^{139}$Gd & J.M. Nitschke & Z. Phys. A &\cite{1983Nit01}& FE & Berkeley & USA &1983 \\
$^{140}$Gd & C.J. Lister & Phys. Rev. Lett. &\cite{1985Lis01}& FE & Daresbury & UK &1985 \\
$^{141}$Gd & N. Redon & Z. Phys. A &\cite{1986Red01}& FE & Grenoble & France &1986 \\
$^{142}$Gd & S. Lunardi & Z. Phys. A &\cite{1986Lun01}& FE & Juelich & Germany &1986 \\
$^{143}$Gd & K.L. Kosanke & Nucl. Instrum. Meth. &\cite{1975Kos01}& LP & Michigan State & USA &1975 \\
$^{144}$Gd & K.A. Keller & Radiochim. Acta &\cite{1968Kel01}& LP & Karlsruhe & Germany &1968 \\
$^{145}$Gd & J.R. Grover & Phys. Rev. &\cite{1959Gro01}& LP & Brookhaven & USA &1959 \\
$^{146}$Gd & G.M. Gorodinskii & Bull. Acad. Sci. USSR &\cite{1957Gor01}& SP & Dubna & Russia &1957 \\
$^{147}$Gd & V.S. Shirley & Nucl. Phys. &\cite{1957Shi01}& LP & Berkeley & USA &1957 \\
$^{148}$Gd & J.O. Rasmussen & Phys. Rev. &\cite{1953Ras01}& LP & Berkeley & USA &1953 \\
$^{149}$Gd & R.W. Hoff & Phys. Rev. &\cite{1951Hof01}& LP & Berkeley & USA &1951 \\
$^{150}$Gd & J.O. Rasmussen & Phys. Rev. &\cite{1953Ras01}& LP & Berkeley & USA &1953 \\
$^{151}$Gd & R.E. Hein & Phys. Rev. &\cite{1950Hei01}& LP & Ames & USA &1950 \\
$^{152}$Gd & A.J. Dempster & Phys. Rev. &\cite{1938Dem01}& MS & Chicago & USA &1938 \\
$^{153}$Gd & M.G. Inghram & Phys. Rev. &\cite{1947Ing02}& NC & Argonne & USA &1947 \\
$^{154}$Gd & A.J. Dempster & Phys. Rev. &\cite{1938Dem01}& MS & Chicago & USA &1938 \\
$^{155}$Gd & F.W. Aston & Nature &\cite{1933Ast01}& MS & Cambridge & UK &1933 \\
$^{156}$Gd & F.W. Aston & Nature &\cite{1933Ast01}& MS & Cambridge & UK &1933 \\
$^{157}$Gd & F.W. Aston & Nature &\cite{1933Ast01}& MS & Cambridge & UK &1933 \\
$^{158}$Gd & F.W. Aston & Nature &\cite{1933Ast01}& MS & Cambridge & UK &1933 \\
$^{159}$Gd & F.D.S. Butement & Phys. Rev. &\cite{1949But01}& NC & Harwell & UK &1949 \\
$^{160}$Gd & F.W. Aston & Nature &\cite{1933Ast01}& MS & Cambridge & UK &1933 \\
$^{161}$Gd & F.D.S. Butement & Phys. Rev. &\cite{1949But01}& LP & Harwell & UK &1949 \\
$^{162}$Gd & M.A. Wahlgren & Phys. Rev. &\cite{1967Wah01}& NC & Savannah River & USA &1967 \\
$^{163}$Gd & R.J. Gehrke& Radiochim. Acta &\cite{1982Geh02}& SF & Idaho Falls & USA &1982 \\
$^{164}$Gd & R.C. Greenwood & Radiochim. Acta &\cite{1988Gre01}& SF & Idaho Falls & USA &1988 \\
$^{165}$Gd & S. Ichikawa & Phys. Rev. C &\cite{1998Ich01}& CPF & JAERI & Japan &1998 \\
$^{166}$Gd & S. Ichikawa & Phys. Rev. C &\cite{2005Ich01}& CPF & JAERI & Japan &2005 \\
 & & & & & & & \\
 & & & & & & & \\
$^{135}$Tb & P.J. Woods & Phys. Rev. C &\cite{2004Woo01}& FE & Argonne & USA &2004 \\
$^{136}$Tb  & & & & & & & \\
$^{137}$Tb  & & & & & & & \\
$^{138}$Tb  & & & & & & & \\
$^{139}$Tb & Y. Xie & Eur. Phys. J. A &\cite{1999Xie01}& FE & Lanzhou & China &1999 \\
$^{140}$Tb & P.A. Wilmarth & Z. Phys. A &\cite{1986Wil01}& FE & Berkeley & USA &1986 \\
$^{141}$Tb & P.A. Wilmarth & Z. Phys. A &\cite{1986Wil01}& FE & Berkeley & USA &1986 \\
$^{142}$Tb & R.B. Firestone & Phys. Rev. C &\cite{1991Fir01}& FE & Berkeley & USA &1991 \\
$^{143}$Tb & T. Ollivier & Z. Phys. A &\cite{1985Oll01}& FE & Grenoble & France &1985 \\
$^{144}$Tb & D.C. Sousa & Phys. Rev. C &\cite{1982Sou01}& FE & Texas A\&M & USA &1982 \\
$^{145}$Tb & G.D.Alkhazov & Acta Phys. Pol. B &\cite{1981Alk02}& SP & Leningrad & Russia &1981 \\
$^{146}$Tb & E. Newman & Phys. Rev. C &\cite{1974New02}& FE & Oak Ridge & USA &1974 \\
$^{147}$Tb & Y.Y. Chu & Phys. Rev. &\cite{1969Chu01}& SP & Brookhaven & USA &1969 \\
$^{148}$Tb & K.S. Toth & J. Inorg. Nucl. Chem. &\cite{1960Tot01}& FE & Berkeley & USA &1960 \\
$^{149}$Tb & J.O. Rasmussen & Phys. Rev. &\cite{1950Ras01}& LP & Berkeley & USA &1950 \\
$^{150}$Tb & K.S. Toth & Phys. Rev. &\cite{1959Tot01}& LP & Uppsala & Sweden &1959 \\
$^{151}$Tb & J.O. Rasmussen & Phys. Rev. &\cite{1953Ras01}& LP & Berkeley & USA &1953 \\
$^{152}$Tb & K.S. Toth & Phys. Rev. &\cite{1959Tot02}& LP & Berkeley & USA &1959 \\
$^{153}$Tb & J.W. Mihelich & Phys. Rev. &\cite{1957Mih01}& LP & Oak Ridge & USA &1957 \\
$^{154}$Tb & G. Wilkinson & Phys. Rev. &\cite{1950Wil03}& LP & Berkeley & USA &1950 \\
$^{155}$Tb & J.W. Mihelich & Phys. Rev. &\cite{1957Mih01}& LP & Oak Ridge & USA &1957 \\
$^{156}$Tb & G. Wilkinson & Phys. Rev. &\cite{1950Wil03}& LP & Berkeley & USA &1950 \\
$^{157}$Tb & R.A. Naumann & J. Inorg. Nucl. Chem. &\cite{1960Nau01}& LP & Berkeley & USA &1960 \\
$^{158}$Tb & C.L. Hammer & Phys. Rev. &\cite{1957Ham01}& PN & Ames & USA &1957 \\
$^{159}$Tb & F.W. Aston & Nature &\cite{1933Ast01}& MS & Cambridge & UK &1933 \\
$^{160}$Tb & W. Bothe & Naturwiss. &\cite{1943Bot01}& NC & Heidelberg & Germany &1943 \\
$^{161}$Tb & F.D.S. Butement & Phys. Rev. &\cite{1949But01}& LP & Harwell & UK &1949 \\
$^{162}$Tb & V.T. Schneider & Radiochim. Acta &\cite{1965Sch01}& LP & Karlsruhe & Germany &1965 \\
$^{163}$Tb & L. Funke & Nucl. Phys. &\cite{1966Fun01}& LP & Jena & Germany &1966 \\
$^{164}$Tb & E. Monnand & J. Phys. (Paris) &\cite{1968Mon01}& LP & Grenoble & France &1968 \\
$^{165}$Tb & R.C. Greenwood & Phys. Rev. C &\cite{1983Gre01}& SF & Idaho Falls & USA &1983 \\
$^{166}$Tb & S. Ichikawa & Nucl. Instrum. Meth. A &\cite{1996Ich01}& CPF & JAERI & Japan &1996 \\
$^{167}$Tb & M. Asai & Phys. Rev. C &\cite{1999Asa01}& CPF & JAERI & Japan &1999 \\
$^{168}$Tb & M. Asai & Phys. Rev. C &\cite{1999Asa01}& CPF & JAERI & Japan &1999 \\

 & & & & & & & \\
 & & & & & & & \\

 & & & & & & & \\
 & & & & & & & \\

 & & & & & & & \\
 & & & & & & & \\

 \\
\end{longtable}

\end{document}